\documentclass[prd,twocolumn,superscriptaddress,amsfonts,amssymb,amsmath,showpacs]{revtex4-2}
\usepackage{amssymb}
\usepackage{bm}
\usepackage{siunitx}
\usepackage{amsfonts}
\usepackage{latexsym}
\usepackage[latin1]{inputenc}
\usepackage{graphicx}
\usepackage{amsmath}
\usepackage{mathtools}
\usepackage{palatino}
\usepackage{mathpazo}
\usepackage{textcomp}
\usepackage{float}
\usepackage{booktabs}
\usepackage{caption}
\usepackage{subcaption}
\usepackage{dcolumn}
\usepackage{ragged2e}
\usepackage{hyperref}
\hypersetup{colorlinks,citecolor=blue}
\hypersetup{colorlinks=true,linkcolor=red,filecolor=magenta, urlcolor=blue}
\usepackage{mathrsfs} 
\usepackage[many]{tcolorbox} 
\newtcolorbox{boxB}{
    fontupper = \bf, % font color
    boxrule = 1.5pt,
     width = 18cm,
    colframe = black,
    colback = white,
    rounded corners,
    arc = 5pt   % corners roundness
    }

\usepackage{amsmath}
\usepackage{xcolor}
\usepackage{orcidlink}
\usepackage{epsfig}
\usepackage{commath}
\usepackage{tabularx}
\usepackage{multirow}
\usepackage{mathtools}
\newcommand{\be}{\begin{equation}}
	\newcommand{\ee}{\end{equation}}
\newcommand{\bea}{\begin{eqnarray}}
	\newcommand{\eea}{\end{eqnarray}}
\newcommand{\beas}{\begin{eqnarray*}}
	\newcommand{\eeas}{\end{eqnarray*}}
\newcommand{\x}{\mathrm{x}}
\newcommand{\T}{\mathrm{T}}
\newcommand{\I}{\mathrm{I}}
\newcommand{\J}{\mathrm{J}}
\newcommand{\B}{\mathrm{B}}
\newcommand{\R}{\mathrm{R}}
\newcommand{\M}{\mathrm{M}}
\newcommand{\N}{\mathrm{N}}
\newcommand{\G}{\mathrm{G}}
\newcommand{\C}{\mathrm{C}}
\newcommand{\BM}{\mathrm{BM}}
\newcommand{\CFL}{\mathrm{CFL}}
\newcommand{\OM}{\mathrm{OM}}
\newcommand{\DM}{\mathrm{DM}}
\newcommand{\tot}{\mathrm{tot}}
\newcommand{\Ms}{\mathrm{M}_\odot}

\allowdisplaybreaks[1]
\begin{document} 
	\color{black} %% For one column
\title{Dark Matter Admixed Quark Stars: A Relativistic Two-Fluid Approach}

\author{N. Priyobarta\orcidlink{0009-0000-3417-9764}}
\email{priyo.naoremcha@gmail.com}
\affiliation{Department of Mathematics, Birla Institute of Technology and Science, Pilani, Hyderabad Campus, Jawahar Nagar, Kapra Mandal, Medchal District, Telangana-500078, India.}
\author{S. K. Maurya\orcidlink{0000-0003-4089-3651}}
\email{sunil@unizwa.edu.com}
\affiliation{Department of Mathematics and Physical Science, College of Arts and Science, University of Nizwa, Sultanate of Oman}

\author{Ksh. Newton Singh\orcidlink{0000-0001-9778-4101}}
\email{ntnphy@gmail.com}
\affiliation{Department of Physics \& Astrophysics, University of Delhi, Delhi-110007, India}

\author{B. Mishra\orcidlink{0000-0001-5527-3565}}
\email{bivu@hyderabad.bits-pilani.ac.in}
\affiliation{Department of Mathematics, Birla Institute of Technology and Science, Pilani, Hyderabad Campus, Jawahar Nagar, Kapra Mandal, Medchal District, Telangana-500078, India.}

\begin{abstract}
   
In this paper, we examine the stellar properties of quark stars containing dark matter, focusing on both non-rotating and slowly rotating configurations. By employing a two-fluid framework, we formulate the generalized Tolman--Oppenheimer--Volkoff equations, treating dark matter and ordinary matter as independent perfect fluids that interact solely through gravitational forces. We model ordinary matter using the color-flavor-locked and the MIT bag model equation of state, while for dark matter, we apply a self-interacting bosonic condensate along with fermionic equation of state. By considering a dark matter fraction of $f_\x=5\%$ and varying the bag constant for ordinary matter, we investigate how dark matter accumulation affects global stellar features such as maximum gravitational mass, radius, and dimensionless tidal deformability. Further, we extend our analysis to first-order rotational effects, calculating the frame-dragging equation and the influence of moments of inertia on the two-fluid system. We also explore universal relations, particularly the connection between rotation and tidal effects, to understand how the inclusion of dark matter affects the relationship between rotational and the tidal deformability. Our findings indicate that a dark matter fraction of $f_\x=5\%$ can lead to notable deviations from the traditional single-fluid quark star characteristics in stellar radius, gravitational mass and tidal deformability. Moreover, the response to rotational and tidal forces remains consistent even in the inclusion of the dark matter component. Finally, we compare our theoretical results with current observational data from GW events and the NICER mission, thereby demonstrating that our developed two-fluid models are consistent with these observations.\\


\end{abstract}		
%\pacs{04.50+h}
\maketitle

\section{Introduction}\label{Sec:I}

For the past century, compact stars have been acknowledged as unique environments for exploring fundamental physics in the realm of extreme gravity. This area of research emerged from the analytic study of a static neutron star (NS) by  Oppenheimer and Volkoff \cite{Oppenheimer_1939} and Tolman \cite{Tolman_1939} in 1939. Since then, numerous studies have calculated the equilibrium structures of black holes (BHs), NSs, cold white dwarfs (WDs), quark stars and supermassive stars in non-rotating configurations. However, these models typically assume a specific baryonic or quark equation of state (EoS), whereas the exact EoS of dense matter in the interiors of compact objects remains a mystery \cite{Lattimer_2001,Lattimer_2004}. This uncertainty motivates the theoretical framework to consider additional components, such as dark matter, that interact with ordinary matter only gravitationally \cite{Goldman_1989,Kouvaris_2010,Kain_2021}. Although the characteristics of dark matter particles are not yet fully understood, NS and strange quark stars provide unique natural laboratories for exploring extreme regimes of density, gravity, and composition.

Recently, researchers \cite{Grippa_2024} have discussed the possibility that these compact objects could gather dark matter from their surroundings. The possibility of this accumulation of dark matter through either inheritance during the initial formation process in dark matter-rich environments or capture over astrophysical timescales through ambient scattering of dark matter particles with ordinary matter \cite{Brito_2015,Guver_2014,Bertone_2008}. When dark matter particles are captured, they gradually lose kinetic energy as they interact with the surrounding stellar medium, becoming gravitationally bound to the compact objects. The configuration of this dark matter affected the mass of the particles and their self-interactions. Consequently, there are two primary categories of dark matter admixed compact stars: those that have a central dark matter core $\left(\gtrsim1\,\text{GeV}\right)$ and those that possess an extended dark matter halo $\left(\sim 100\,\text{MeV}\right)$ surrounding them \cite{Ivanytskyi_2020,Aprile_2023,Rutherford_2025}. With heavier dark matter particles, the dark matter core diminishes the radius, tidal deformability and maximum mass, while lighter particles extend into a halo, increasing the tidal deformability and radius \cite{Bell_2020,Liu_2024,Koehn_2024}.

Inclusion of non-gravitational couplings between ordinary matter and dark matter effectively reduces the system to a single fluid. In this work, we instead neglect such interactions altogether. This approximation is well-motivated as experimental constraints imply that any non-gravitational coupling is extremely weak, and the properties of dark matter, like self-interaction strength and mass, remain poorly known \cite{Das_2022}. Under this assumption, dark matter and ordinary matter constitute the two-fluid system interacting only through gravity \cite{Leung_2011}. In the study of non-rotating stellar configurations, numerous researchers have extensively explored the inclusion of dark matter particles, both bosonic and fermionic, in the context of NSs (see Refs. \cite{Das_2020,Sanchis_2022,Sousa_1998,Brito_2016}). These findings indicate that incorporating dark matter can significantly modify stellar characteristics, including radius, mass, and tidal deformability \cite{Ellis_2018,Kumar_2026}.

While much of the existing literature focuses on static configurations, the inclusion of rotation provides additional diagnostics of the internal structure and composition \cite{Rios_2025,Mourelle_2024,Kain_2021,Kumar_2026}.
 Following the pioneering research study by Hartle and Thorne \cite{Hartle_1967,Hartle_1968}, one can incorporate the influence of rotation as perturbations to the spherically symmetric line element of a static star to estimate the line element with slow, rigid rotation. Studies \cite{Meltzer_1966,Wheeler_1966} have suggested that the damping of radial oscillations of a NS may depend on its rotation. Further, the studies have predicted such coupling between the rotation and radial, as well as quadrupole, modes of oscillation that radiate gravitational energy via the radial mode. It is to be noted that the gravitational collapse of supermassive stars can be countered by their rotation just before the beginning of nuclear fusion at their core \cite{Fowler_1966, Meltzer_1966}. In the slowly rotating approximation of a two-fluid stellar system, rotation is considered only the linear-order perturbation to the static or non-rotating background. This spinning induces a frame-dragging effect in the system, leading to varying angular velocities of the local inertial frames within the stellar system. As a result, the two fluids may rotate in a rigid manner while maintaining distinct angular velocities, and the subsequent frame-dragging function establishes the overall moment of inertia of the stellar system \cite{Cipriani_2025,Andersson_2001}. As dark matter alters the properties of the stellar system, these compact objects will be relevant for probing tidal deformability. In a binary system, the gravitational field of a companion star causes each star to experience quadrupole deformation. The tidal response is characterized by the dimensionless tidal deformability, a parameter that is critically informed by gravitational-wave (GW) observations. Notable events such as GW170817 \cite{Abbott_2017_18} and GW190425 \cite{Abbott_2020} provide direct constraints on this deformability, influencing our understanding of the EoS for dense matter in compact stars. The inclusion of dark matter, whether concentrated in a core or distributed in an extended halo, has the potential to significantly influence the internal density profiles of stellar structures \cite{Leung_2022}. Recent studies \cite{Ellis_2018,Sennett_2017,Maselli_2017} demonstrate that even with a minute amount of dark matter, it can greatly affect the conclusions about the exclusion of NS EoS. This dimensionless tidal deformability, along with the moments of inertia, plays a vital role in modelling the spin evolution of pulsars and in developing EoS-insensitive relationships, known as universal relations \cite{Yagi_2013_023,Yagi_2013_365}.
 These works are extensively done for both bosonic and fermionic dark matter with realistic EoS, while the analysis within the fully gravitationally coupled two-fluid formalism, considering quark stars, remains comparatively less developed. The advancement of multi-messenger astronomy has provided a robust empirical framework for testing these two-fluid systems. Precise mass-radius measurement for the NICER missions such as PSR J0740$+$6620 ($2.072\pm 0.07\ \Ms$ with radius $12.35\pm 0.75$ km), PSR J0437$-$4715 ($1.418\pm0.037 \ \Ms$ with radius $11.36^{+0.95}_{-0.63}$ km) and PSR J0030$+$0451 ($1.44^{+0.15}_{-0.14}\ \Ms$ with radius $13.02^{+1.24}_{-1.06}$ km) \cite{Miller_2021,Riley_2021,Miller_2019,Choudhury_2024} alongside GW events constrain from GW190425 and GW170817 binary merger have become indispensable for narrowing the vast parameter space of dark matter microphysics. On top of the constraints from the  NICER measurements and GW data, the compact object within the supernova remnant HESS J1731$-$347 offers a significant constraint on the mass-radius relationship for low-mass stellar remnants. X-ray spectral modeling combined with a Gaia-based distance estimate inferred $\M=0.77^{+0.20}_{-0.17}\,\Ms$ and radius $\R=10.4^{+0.86}_{-0.78}$ km \cite{Doroshenko_2022} and also included both precisely measured pulsar masses and candidate compact objects in the NS-BH mass gap, particularly, PSR J0384$+$0432 with measured mass $2.01\pm0.04\, \Ms$ \cite{Antoniadis_2013}  and the secondary component of GW190814 was inferred to have a mass within $2.50-2.67\, \Ms$ \cite{Abbott_2020}. This study aims to derive the gravitational field equations describing dark matter admixed quark stars, focusing on configurations in static equilibrium and slow rotation. We will also analyze the consequent effects of dark matter on the macroscopic structure and observable characteristics of these stellar objects.

 In this work, we consider four EoS, namely the Unified MIT bag model (BM) \cite{Holdom_2018,Zhang_2021}, color--flavor--locked (CFL) \cite{Alford_2005,Alford_2001}, bosonic dark matter \cite{Colpi_1986,Rafiei_2022}, and fermionic dark matter \cite{Narain_2006,Serot_1992}, to investigate the various stellar properties with a two-fluid system. The EoS for ordinary matter will be based on the first two items, while the last two will represent the EoS for dark matter. The primary aim of employing EoS for these two types of matter is to develop a two-fluid model for both static and slowly rotating stellar systems in the present study. We begin with the non-rotating, spherically symmetric metric and derive the coupled TOV equations for the two-fluid system, in which the dark matter and ordinary matter components interact solely through gravity. We then extend the analysis to the slow-rotation approximation of Hartle and Thorne \cite{Hartle_1967,Hartle_1968}, building on the generalization of the Hartle-Thorne formalism within a two-fluid framework. Within this approach, we derive and solve the corresponding frame-dragging equation for the two-fluid system and finally compute the moment of inertia to quantify the influence of dark matter on the rotational stellar system. We examine the effects of dark matter on the tidal deformability, providing a comprehensive assessment of how dark matter modifies both the rotational and tidal properties of the stellar system. In conclusion, we present the universal relation characterized by the rotational-tidal relationship to evaluate the validity and robustness of the two-fluid stellar system configuration.

The paper is structured as follows: In Section \ref{Sec:II}, we provide a brief overview of two EoS for ordinary matter and two EoS for dark matter. Section \ref{Sec:III} describes the various internal structures of the two-fluid stellar system. The generalized two-fluid TOV equations are established in section \ref{Sec:IIIa}, while the slow-rotation formalism that addresses the frame-dragging equation and moment of inertia is detailed in section \ref{Sec:IIIb}. In section \ref{Sec:IIIc}, we provide the formalism utilized for calculating the tidal deformability of the two-fluid stellar system. Section \ref{Sec:IIId} includes a brief examination of the universal rotational-tidal relationship. The findings and discussions are thoroughly outlined in section \ref{Sec:IV}. Lastly, the conclusions derived from this work are summarized in section \ref{Sec:V}.

\section{Equations of State for Ordinary Matter and Dark Matter}\label{Sec:II}

\subsection{Unified MIT Bag Model}

To characterize ordinary quark matter, we utilize the MIT bag Model EoS, which represents a relativistic gas of deconfined quarks \cite{Chodos_1974_347,Chodos_1974_259,Farhi_1984},
\begin{eqnarray}
 p = \frac{1}{3}\left(\rho - 4\B_g\right),
\end{eqnarray}
where $\B$ is known as the bag constant. In the standard MIT bag model, the stability window corresponds to $\B_g^{1/4}\sim145$--$200~\mathrm{MeV}$ \cite{Farhi_1984}, such values are widely employed in studies of hybrid compact stars and strange quark star \cite{Weissenborn_2011,Ivanov_2005,Fraga_2014}. However, bag constants within this range are generally not sufficiently small to accommodate constraints from recent binary-merger observations within the framework of  GR alone \cite{Gammon_2024}.
 
From the Holdem et al. works \cite{Holdom_2018}, Zhang and Mann \cite{Zhang_2021} show that a single parameter $\tau$, can capture strong interaction effects and yield a quark matter EoS consistent with current GR observations. This parametrization optimizes the analysis by eliminating the need to solve the stellar equations for distinct phases of quark matter, including two-flavour superconducting phases with or without strange quarks (2SC+s and 2SC), the CFL phase, varying color-superconducting gap parameters, and the effects of perturbative Quantum Chromodynamics (pQCD) corrections. Instead, all such effects are effectively incorporated into the single parameter $\tau$. To derive the results, the  expression of the free energy $\mathcal{V}_{\BM}$ of superconducting quark matter in a specific formalism is given as \cite{Alford_2002,Fraga_2001,Alford_2005} 
\begin{eqnarray}
\label{eq:Omega}
    \mathcal{V}_\BM &=& -{\Xi_4\over 4\pi^2}~\mu^4 + {\Xi_4(1-a_4)\over 4 \pi^2}~\mu^4 \nonumber \\
   &&- {\Xi_{2a}\, \Pi^2-\Xi_{2b} \,m^2_s \over \pi^2} ~\mu^2  -{\mu_e^4 \over 12\pi^2}+ \B_g
\end{eqnarray}
where $\Pi$, S$\mu$, $m_s$, $a_4$ and $\mu_e$ are the gap parameter, the average quarks, corrections from the finite strange quark mass, pQCD contribution from one-gluon exchange and electron chemical potentials, respectively. The values of the coefficients of Eq.~\eqref{eq:Omega} are as follows:
\begin{footnotesize}
%\begin{widetext}
\begin{equation}
(\Xi_{4},\,\Xi_{2a},\,\Xi_{2b})=
\begin{cases}
\left(\left[\left(\dfrac{1}{3}\right)^{4/3}
+\left(\dfrac{2}{3}\right)^{4/3}\right]^{-3}, \,1,\,0\right)
& \text{2SC phase},\\[4.0ex]
\left(3,1,{3\over4}\right)
& \text{2SC+s phase},
\\[2.0ex]
\left(3,3,{3\over4}\right)
& \text{CFL phase},
\end{cases}
\end{equation}
%\end{widetext}
\end{footnotesize}
for the various possible QM phases. Further, using the thermodynamic relations
\begin{eqnarray}
p_{_\BM} &&=-\mathcal{V}_{_\BM},~~\,
n_q=-\frac{\partial \Omega_{_\BM}}{\partial \mu},~~
n_e=-\frac{\partial \mathcal{V}_{_\BM}}{\partial \mu_e},\nonumber \\
\mbox{and}\quad && \rho_{_\BM}=\mathcal{V}_{_\BM}+n_q\mu+n_e ,
\end{eqnarray}
along with the reparameterization
\begin{equation}
\tau=\frac{\Xi_{2a}\,\Delta^{2}-\Xi_{2b}\,m_s^{2}}{\sqrt{\Xi_{4}\,a_{4}}},
\end{equation}
we obtain
\begin{eqnarray}
n_q &=& \frac{\Xi_{4}\,a_{4}}{\pi^{2}}\mu^{3}+\frac{\tau \sqrt{\Xi_{4}\,a_{4}}}{\pi^{2}}~2\mu~,\\
n_e &=& \frac{\mu_e^{3}}{3\pi^{2}}~,\\
\rho_{_\BM} &=& \frac{3\,\Xi_{4}\,a_{4}}{4\pi^{2}}\,\mu^{4}
+\frac{\mu_e^{4}}{4\pi^{2}}+\B_g+\frac{\tau \sqrt{\Xi_{4}\,a_{4}}}{\pi^{2}}\,\mu^{2}~.\label{eq:RhoBM}
\end{eqnarray}
By combining Eq.~\eqref{eq:RhoBM} with Eq.~\eqref{eq:Omega}, we can express the pressure in terms of the MIT bag constant,  the unified interaction strength, and the mass density, ($\rho_{_\BM}$) as 
\begin{eqnarray}
\hspace{-2cm} p_{_\BM} &=&\,\frac{1}{3}\left(\rho_{_\BM} -4\B_g\right) +\frac{4\tau^{2}}{9\pi^{2}} \nonumber \\
&& \left(\mathrm{sgn}(\tau)\sqrt{1+\frac{3\pi^{2} \left(\rho_{_\BM}-\B_g\right)}{\tau^{2}}}-1\right).
\end{eqnarray}
Here, $\mathrm{sgn}(\tau)$ represent the sign of the $\tau$. Gammon et al. \cite{Gammon_2024} investigate both scenarios that involve both negative and positive signs, discovering that an increased value of $\tau$ typically strengthens with the higher masses and radii of quark stars, while a significantly negative $\tau$ produces the contrary effect, counteracting the impact of non-zero coupling on the terms related to higher curvature gravity. In our research, we will concentrate exclusively on the positive value, specifically $\tau = 0.5$.

\subsection{Color-Flavor-Locked }

We also examine an alternative EoS for CFL quark matter, derived from the MIT bag model. In the CFL phase, the thermodynamic potential for electrically and color charge neutral CFL quark matter is expressed as \cite{Alford_2001}:
\begin{eqnarray}\label{eq:OmegaCFL}
\hspace{-2cm} \mathcal{V}_{_\CFL} &=& \frac{6}{\pi^2} \int_{0}^{P} (p - \mu) p^2 \, dp- \frac{3}{\pi^2} \Delta^2 \mu^2 + \B_g \nonumber \\
&&+ \frac{3}{\pi^2} \int_{0}^{P} \left[\sqrt{(p^2 + m_s^2)} - \mu \right] p^2 \, dp 
\end{eqnarray}
where, $\Delta$ denotes the color superconducting gap parameter of CFL phase, $\mu$ is the quark chemical potential and $P$  is the Fermi momentum given as 
\begin{eqnarray}
P= 2\mu -\sqrt{\mu^2 + {m^2_s\over 3}} \sim \mu -{m^2_s \over 6 \mu}~.
\end{eqnarray}
Lugones et. al. \cite{Lugones_2002} derive the pressure and energy density at zero temperature, given as, respectively,
\begin{eqnarray}
p_{_\CFL} = -\mathcal{V}_{_\CFL},~~~ \mbox{and}~~~ \rho_{_\CFL} = \sum_i \mu_i n_i +\mathcal{V}_{_\CFL} -p_{_\CFL} \,.
\end{eqnarray}
Due to the challenges in obtaining an exact analytical expression for the EoS when the strange quark mass $m_s\neq0$, an alternative approximate approach can be employed for $m_s\xrightarrow{}0$. In this framework, a simplified EoS similar to that of the MIT bag model can be developed, incorporating an additional term $\Delta^2$ contribution from \eqref{eq:OmegaCFL}, the CFL pairing as ~~$\rho =3p + 4\B_g - 6\Delta^2\mu^2 /\pi^2$.

By expanding the thermodynamic potential in terms of the strange quark mass and keeping only the relevant terms, one can derive approximate expressions for the pressure  and energy density in the CFL phase \cite{Lugones_2002} as,
\begin{eqnarray}
p_{_\CFL} &=& {3\mu^4 \over 4 \pi^2} + {9\Lambda \mu^2 \over 2 \pi^2}-\B_g~~ \mbox{and}~~ \nonumber\\ 
\rho_{_\CFL} &=& {9\mu^4 \over 4 \pi^2} + {9\Lambda \mu^2 \over 2 \pi^2}-\B_g~~,
\end{eqnarray}
where, $\Lambda = -m^2_s/ 6 +\Delta^2 /3$ and the finally we obtained an expression $p_{_\CFL}$ as an explicit function of the energy density $\rho_{_\CFL}$ as
\begin{eqnarray}
\rho_{_\CFL} &=& 3p_{_\CFL} + 4\B_g \nonumber \\ 
&&-{9\Lambda \over  \pi^2}\left(\sqrt{{4\pi^2(p_{_\CFL} +\B_g) \over 3}}- 3\Lambda\right)\label{eq:CFLeq}
\end{eqnarray}
In our article, we will consider $m_s \sim 150\, \text{MeV}$ and $\Delta \sim 100\, \text{MeV}$ as in the Ref. \cite{Lugones_2002} which is  As mentioned in the Ref. \cite{Farhi_1984}, the necessary condition for MIT-based EoS, the bag constant, is always greater than $57\, \text{MeV/fm}^3 $. We vary the values of the bag constant from within the range of $60-75\,\text{MeV/fm}^3$.

\subsection{Bosonic Dark Matter}
We consider self-interacting bosonic DM described by the Lagrangian \cite{Colpi_1986}
\begin{eqnarray}
    \mathcal{L}={1\over2} \partial_\mu\psi^* \partial^\mu \psi -{1\over 2}m_{\x} \left(\psi^*\psi\right) -{1\over 4}\eta \left(\psi^*\psi\right)^2
\end{eqnarray}
where $m_\x$ represents the mass of DM particles, $\eta$ is a dimensionless coupling constant, and $\psi$ is the complex scalar field that forms as a stable Bose-Einstein condensate at low temperature. Following the computation as provided in Refs. \cite{Rafiei_2022,Shakeri_2024}, one can obtain the total pressure as

\begin{equation}
    p_{\x} = {m^4_\x \over9\eta}\left(\sqrt{{3\eta\,\rho_\x\over m^4_\x}+1}-1\right)^2~~,
\end{equation}
where $\rho_\x$ is the DM energy density. In the nonrelativistic limit, the EoS behaves similarly to a polytropic model, with pressure $p_\x $ scaling like  $\rho_\x^2$. As  $\rho_\x$ increases, the system transitions towards the relativistic regime characterized by $p_\x = \rho_\x/3$. Following the parameterization outlined in Ref. \cite{Cipriani_2025}, we set the mass of the DM particle $m_\x = 1 \,\text{GeV}$ and choose $\eta = 24\pi $. This configuration reflects a framework for strongly self-interacting bosonic DM, which is critical to achieving a sufficiently compact stellar structure capable of forming a central DM core within the stellar system \cite{Shakeri_2024}. 
%%%%%%%%%%%%%%%%%%%%%%%%%%%%%%%%%%%%%%%%%%%%%%%%%%%%%%%%%%%%%%
\subsection{Fermionic Dark Matter}

Finally, we explore a two-body interaction model for fermions, utilizing the lowest order approximation. This approach reveals that the interaction energy density is proportional to the square of the number density, $n$S. One can express this relationship mathematically as $\rho_{\rm int} = n^2/m^2_{\mathrm{I}}$, where  $m^2_{\rm I}$ characterizes the energy scale of the interaction. Additionally, the pressure can be articulated in Ref.\cite{Narain_2006} as
\begin{equation}
p_{\rm int} = -\left.\frac{\partial E}{\partial V}\right|_{N,T=0}
=n^{2}~\frac{\partial (\rho_{\rm int}/n)}{\partial n}
= \frac{n^{2}}{m_{I}^{2}} ~,
\end{equation}
where the pressure and energy density include an extra term of $n^{2}/m_{I}^{2}$. The interaction should be repulsive, so that an increase in number density leads to a higher in pressure and energy density. The scale $m_{I}$ can be interpreted as the vacuum expectation value of the Higgs field associated with this interaction \cite{Narain_2006}. For strong interactions, the typical mass scale is approximately $m_I \sim 100~\mathrm{MeV}$. For a more accurate representation, it is essential to incorporate an attractive scalar interaction alongside the repulsive vector interaction. This approach is frequently applied to both baryonic and quark matter stars \cite{Hanauske_2001,Serot_1992}. In dimensionless forms, the pressure and energy density are written, respectively, as
\begin{eqnarray}
p_{f}&=& \frac{1}{24\pi^2}\left[z(2z^2-3)\sqrt{1+z^2} +3\sinh^{-1}z \right]\nonumber \\
&& +\left(\frac{1}{3\pi^2}\right)^2y^2 z^6 \quad  \\
\rho_{f} &=& \frac{1}{8\pi^2}\left[z(2z^2+1)\sqrt{1+z^2}-\sinh^{-1}z \right]\nonumber \\
&&+\left(\frac{1}{3\pi^2}\right)^2y^2 z^6~,
\end{eqnarray}
where $z$ denotes the dimensionless Fermi momentum, while $y$ is defined as the mass ratio of the fermion to the interaction mass, expressed as $y = m_f/m_{\rm I}$. For a realistic NS, neutrons experience strong interactions with an interaction mass of approximately $m_I \sim 100~\mathrm{MeV}$. Given that the neutron mass is around $m_f \sim 1~\mathrm{GeV}$, we can conclude that $y \sim 10$. In our case, we will consider  $m_f \sim 1.5~\mathrm{GeV}$ such that we will get a radius $\leq 10$ km \cite{Narain_2006}.

\section{Interior Framework of two-fluid stellar system}\label{Sec:III}
\subsection{Non-Rotating Structure Equations}\label{Sec:IIIa}
For a non-rotating stellar configuration, we adopt a static and spherically symmetric spacetime of the form
\begin{equation}
\label{eqn:metric}
ds^2 =
-e^{2\varphi(r)}dt^2
+e^{2\lambda(r)}dr^2
+r^2\left(d\theta^2+\sin^2\theta\,d\phi^2\right),
\end{equation}
where the metric potential $\varphi(r)$ and $\lambda(r)$ that characterize the temporal and radial components of the spacetime, respectively. We further define the
mass function $m(r)$ as $e^{-2\lambda(r)}=1-{2m(r)}/{r}$,
 so that the metric reduces consistently to the Schwarzschild solution outside the matter distribution.

For the ideal two fluids, hydrostatic equilibrium can be described by neglecting interactions among them, in which the total energy-momentum tensor, provided the continuity equation holds,
\begin{equation}
\nabla_{\mu} \T^{\nu\mu}_{\tot} =\sum_i\nabla_{\mu} \T^{\nu\mu}_i=\nabla_{\mu}\T^{\nu\mu}_{_\OM}+\nabla_{\mu}\T^{\nu\mu}_{_\DM}=0~~,
\end{equation}
where $i$ are ordinary matter (OM) and dark matter (DM), respectively. Finally, for the static configuration, the generalized TOV equations for the two-fluid stellar system are given \cite{Oppenheimer_1939,Tolman_1939} as
\begin{eqnarray}
    \label{eq:TOV}
    {dp_i\over dr} &=& -(\rho_i +p_i)\,{m_{\tot}+4\pi r^3 p_\tot\over r(r-2m_\tot)} \quad \\
     {dm_i\over dr} &=& 4\pi r^2 \rho_i \\
     {d\varphi  \over dr} &=& {m_\tot + 4\pi r^3 p_\tot \over r(r-2m_\tot)}~~.
\end{eqnarray}
Here, $m_{\tot}= m_{_\OM} +m_{_\DM}$ and $p_\tot= p_{_\OM}+p_{_\DM}$. As mentioned earlier, we investigate all possible combinations obtained with two EoSs for OM and two EoS for DM. The TOV equations are integrated outward from the stellar core, where we establish boundary conditions at the centre. Specifically, we need to define the central pressure for OM and DM,  $p_{_\OM}(0)$ and $p_{_\DM}(0)$ respectively, along with the condition $m_{_\tot}(0) = 0$. To effectively integrate the structure equations, we must set the central values for both the EoS components. Various papers defined dark matter fraction in two approaches (see Refs.~\cite{Lopes_2018,Cipriani_2025,Das_2022,Lopes_2023,Marzola_2025}). For our case, the $\DM$, the central pressure will be provided with the $\DM$ fraction, $f_\x$, which is defined as
\begin{eqnarray}
    f_\x = {p_{_\DM}(0) \over p_{_\BM}(0)+p_{_\DM}(0)} ~~.
\end{eqnarray}
The coupled TOV equations are solved numerically using the Runge-Kutta 45 (RK45) method implemented in the
\texttt{scipy.integrate} module of Python. Since the OM and DM are considered as a separate fluid, their respective radii are determined individually from the conditions: $p_{_\OM}(\R_{_\OM}) \simeq 0$ and $p_{_\DM}(\R_{_\DM}) \simeq 0$.  Whereas the stellar surface is identified by the outmost fluid boundary $\R= \max\left(\R_{_\OM},\R_{_\DM}\right)$ and the corresponding gravitational mass is obtained from the mass function as $M=m_{\mathrm{tot}}(\R).$ Depending upon the values of $f_\x$, one can get various deviations of the stellar properties as shown in the Refs. \cite{Cipriani_2025,Lopes_2018,Das_2022}. In order to understand the deviation of the properties of masses and radii, we provide the $\M-\R$ plots for the case of BM admixed bosonic DM and CFL admixed fermionic DM. Since OM is directly observable, we consider the observable stellar radius appearing in the $\M-\R$ relations, as in Refs.~\cite{Lopes_2018,Lopes_2023}.

\begin{figure}[ht]
\centering
\includegraphics[width=0.95\linewidth]{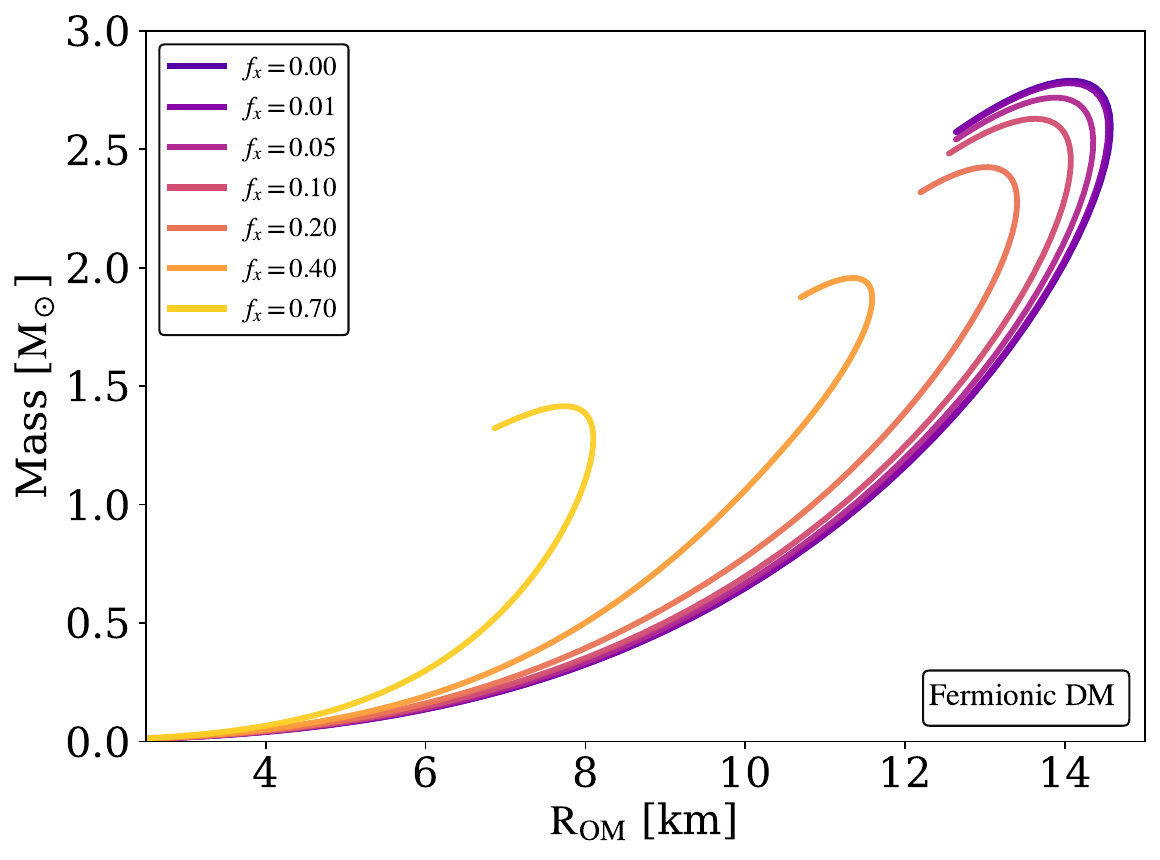}
\vspace{0.3cm}
\includegraphics[width=0.95\linewidth]{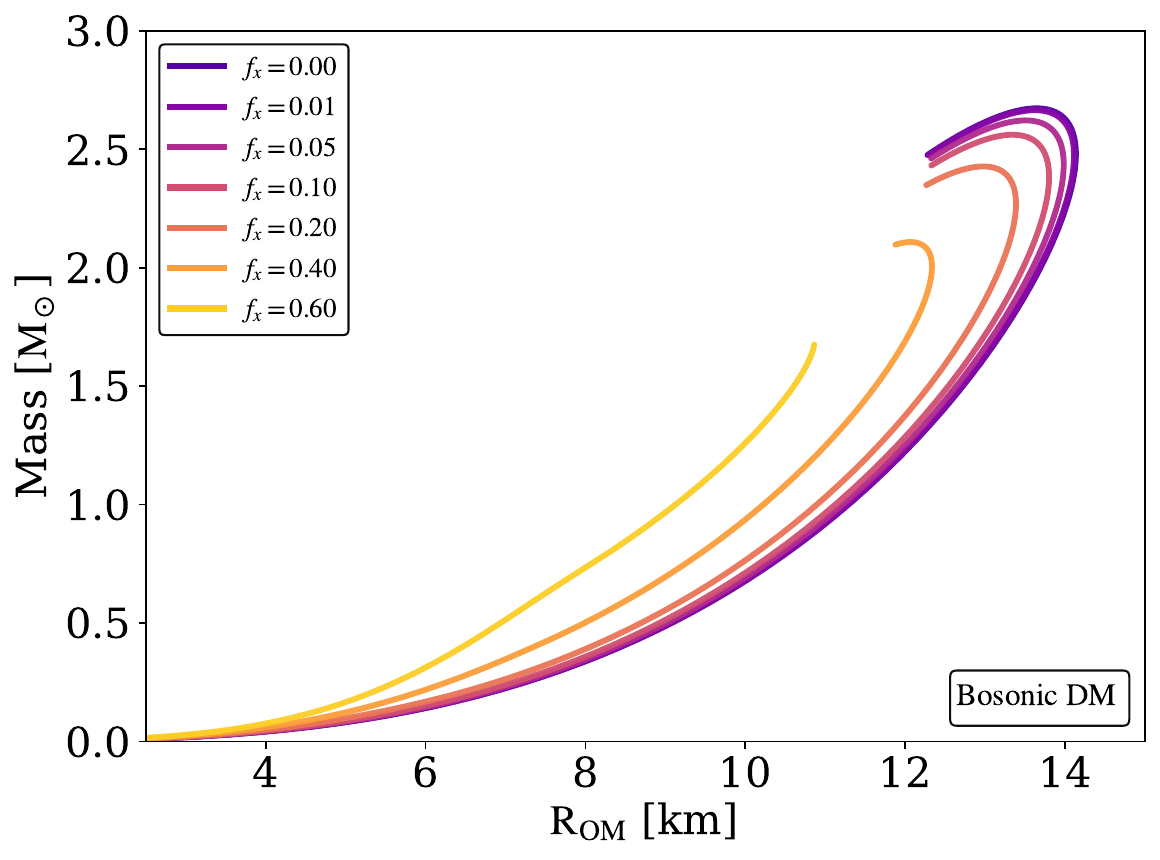}
\caption{\justifying $\M-\R$ relations for quark stars admixed with DM for different DM fractions $f_\x$. The upper panel shows the results for the BM EoS with fermionic DM, while the lower panel corresponds to the CFL EoS with bosonic DM. Different curves denote the DM fractions indicated in the legends. Here, we consider $\B_g= 60\,\mathrm{MeV /fm}^3$ for both BM and CFL and $f_\x=0.0$ corresponds to OM EoS only( without DM EoS)}\label{fig:MR_DM_combined}
\end{figure}

As shown in Fig.~\ref{fig:MR_DM_combined}, increasing the DM fraction, $f_\x$, considerably alters the equilibrium structure of the compact star configurations. For sufficiently large values of $f_\x$, the resulting mass-radius sequences exhibit significant deviations from the standard $\OM$ branch and tend to produce less massive and more compact stellar configurations. In contrast, lower DM fractions maintain the characteristic behavior of strange stars while still capturing the essential effects of DM components. Motivated by this behavior, we adopt $f_\x=5\% $ as a fiducial choice in the present analysis, since it introduces non-negligible DM effects without substantially distorting the underlying stellar structure.

% In this study, we examine the OM EoS, \textit{viz} BM and CFL, admixed with two DM EoS, i.e., bosonic and fermionic. Our goal is to investigate the impact of DM on stellar structure. The OM radius, denoted as $\R_{_\OM} $, is determined by the condition $p_{_\OM}(\R_{_\OM}) = 0$, while the DM radius is found using the condition $p_{_\DM}(\R_{_\DM}) = 0$. The total gravitational mass of the star is assessed at the outermost radius of the configuration. 

\subsection{Slowly Rotating Approximation }\label{Sec:IIIb}

\subsubsection{Frame-Dragging Equation}\label{Sec:IIb1}
Hartle and Thorne introduced the slowly rotating approximation, assuming that the angular velocity is sufficiently small such that the stellar configuration remains nearly spherical \cite{Hartle_1968}. In this approach, rotational effects are treated only as first-order perturbations to the static, spherically symmetric background. This rotation provides the off-diagonal component, $g_{_{t\phi}}$, which corresponds to the frame-dragging of the inertial frame. For a slowly rotating spacetime, the metric can be written as
\begin{eqnarray}  
\label{eqn:rmetric}
&& \hspace{-1cm} ds^2 = -e^{2\varphi\left(r\right)}d t^2 +e^{2\lambda\left(r\right)}dr^2+r^2d\theta^2 \nonumber \\
&& \hspace{2cm} +r^2\sin^2\theta \,\left[ \, d\phi^2 - \omega(r) d t^2\right] ~~~,
\end{eqnarray}
where $\omega(r)$ signifies the local angular velocity of the inertial frame compared to a distant observer, incorporating the effects of rotational frame-dragging. Our assumption takes the four velocity of each fluid as $u^\mu_i = \left[e^{-\varphi},0,0,e^{-\varphi} \Omega_i\right]$, where $\Omega_i$  is the total angular velocity of each fluid \cite{Kumar_2026}. Evaluating the  component of the field equations for the slowly rotating metric by considering only the linear term of $\omega$, we obtain
\begin{eqnarray} 
\label{eq:Gtphi}
\G_{t\phi}=e^{-2\lambda} \sin^2\theta\Bigl[r^2 \omega '' 
+r \omega ' \left(4-r \left[\lambda ' 
+\varphi '\right]\right) \nonumber \\
+2 r \omega  \left(\left[r \varphi ' 
+1\right] \left[\lambda '-\varphi '\right]-r \varphi ''\right)\Bigr] ~.
\end{eqnarray}
On simplifying  $\varphi$ and $\lambda$ using static field equations with Eq. \eqref{eq:Gtphi}, we finally obtain the $t\phi$ component reduces
\begin{eqnarray}
\label{eq:SlowEq}
r^2\omega'' &+& \left[4r-r^2(\varphi' +\lambda')\right]\omega' \nonumber \\
&& =-16 \pi \,e^{2\lambda} r^2 \left[ \sum_i (\rho_i + p_i)(\Omega_i -\omega)\right]
\end{eqnarray}
which can be further reduced to
\begin{eqnarray}\label{eq:frameDrag}
&& \hspace{-1cm} {e^{\varphi+\lambda}\over r^4}\left[{d\over dr}\left(r^4 e^{-(\varphi+\lambda)} \,{d \tilde{\omega}_i\over dr }\right)\right]\nonumber \\
&& \hspace{2cm}  =-16 \pi \,e^{2\lambda}  \left[ \sum_i (\rho_i + p_i)\,\tilde{\omega}_i\right]
\end{eqnarray}
where, we define $\tilde{\omega}_i = \Omega_i -\omega$. By considering the single fluid, one can reduce to the original Hartle-Thorne frame-dragging equation \cite{Hartle_1968}. To solve Eq.~(\ref{eq:frameDrag}), we impose a regularity condition at the centre of the stellar system as
\begin{equation}
\left.\frac{d\omega}{dr}\right|_{r=0} = 0~.
\end{equation}
At the surface of the stellar system, the interior solution must transition seamlessly to the vacuum solution on the exterior. In the outer region, where the source term in Eq.~(\ref{eq:frameDrag}) vanishes, the solution demonstrates a behavior of $\omega \propto 1/r^{3}$.This leads to the following boundary condition that must be satisfied at $r = \R_{_\OM}$ as
\begin{equation}
\frac{d\omega}{dr} + \frac{3}{r}\,\omega = 0~.
\end{equation}

The frame-dragging equation is solved simultaneously with the TOV equations \eqref{eq:TOV} using a shooting method procedure. The integration begins at the centre of the star and proceeds toward the surface, with a trial central value, $\omega(0) = \omega_c$, while applying the regularity condition at $r = 0$. The value of $\omega_c$ is iteratively modified until the surface boundary condition at $r = \R_{_\OM}$ is fulfilled, ensuring a smooth transition to the exterior solution.  In the present work, we restrict our analysis to
co-rotating configurations, for which the angular velocities of the OM and DM components are identical i.e., $\Omega_{_\OM} = \Omega_{_\DM}$.

\subsubsection{Moment of Inertia}\label{Sec:IIIb2}
Once the frame-dragging function $\omega$ is computed, we can calculate the total angular momentum, which is defined via the volume integral of the energy-momentum component $\T^t_\phi$ in the context of GR \cite{Bardeen_1971,Glendenning_1994} as
\begin{eqnarray}
   \J =\int \sqrt{-g}~ ~\T^t_\phi \, dr \,d\theta \,d\phi~~,
\end{eqnarray}
where $\sqrt{-g}$ denotes the determinant of the metric of a slowly rotating metric. For a two-fluid system, $\J_\tot$ is the total angular momentum of the stellar system, which can be generalized as 
\begin{align}
\label{eq:Jtot}
\J_{\mathrm{tot}} = \frac{8\pi}{3} \int_{0}^{R_{_\OM}} 
\left[ \sum_i \tilde{\omega}\,(\rho_i + p_i)\right] 
r^4 e^{\lambda - \varphi} \, dr \, .
\end{align}
Furthermore, one can calculate the important quantity, the moment of inertia of the stellar system and is defined as
\begin{eqnarray}
\I = {\J_\tot \over \Omega_{_\OM}} \,.
\end{eqnarray}
Using Eq. \eqref{eq:Jtot}, we get the expression of the moment of inertia as
\begin{eqnarray}
\label{eq:MoI}
\J_\tot ={8\pi \over3} \int_{0}^{R_{_\OM}} \left[ \sum_i \tilde{\omega}\,{(\rho_i + p_i)\over \Omega_{_\OM}}\right]\, r^4 e^{\lambda-\varphi}\, dr~~~.
\end{eqnarray}
The radial integration will extend up to the $\R_{_\OM}$, which defines the stellar surface of the star.

\subsection{Tidal Deformability}\label{Sec:IIIc}
In a compact binary, each constituent experiences a fluctuating tidal field produced by its companion. The resulting deformation is characterized by the tidal deformability of the object and leaves an observable imprint on the GW phase during inspiral. Consequently, GW
measurements provide a means of constraining the tidal properties of compact stars  \cite{Abbott_2017_18,Abbott_2020}. The tidal deformability for a two-fluid system is computed for both fermionic and bosonic DM \cite{Das_2022,Collier_2022,Diedrichs_2023}. In order to solve tidal deformability for the two-fluid system, we have solved a system of differential equations as provided in the Ref.~\cite{Das_2022}, along with the TOV equations \eqref{eq:TOV} simultaneously,
\begin{eqnarray}
\label{eq:Tidal}
    r{dy(r)\over dr} +y(r)^2 + y(r) X(r) + r^2 Y(r) =0 
 \end{eqnarray}
where,
\begin{eqnarray}
X(r) &=& {r-4\pi r^3\left(\rho_\tot-p_\tot\right) \over r-2 m_\tot}\\
\hspace{-1.2cm} Y(r) &=& {4\pi r \over r-2 m_\tot}\Bigg[5\rho_\tot +9p_\tot +\sum_i \left(\rho_i+p_i\right){\partial\rho_i \over \partial p_i}\nonumber\\
&& -{6\over 4 \pi r^2}\Bigg]-4\left[{m_\tot + 4 \pi r^3 p_\tot \over r(r-2 m_\tot)}\right]^2~~~.
\end{eqnarray}
Here, $y(r)$ depends only on the non-rotating stellar configuration. It should be noted that quark stars possess a finite energy density at the stellar surface. Such a behavior, analogous to a first-order phase transition occurring at fixed pressure, implies that while speed of sound, $ c^2_s=\partial p / \partial \rho$ remains continuous, the quantity $\partial \rho / \partial p$ becomes discontinuous. Consequently, the corresponding expression must be modified accordingly (see Refs.~\cite{Takatsy_2020,Postnikov_2010}) as 
\begin{equation}
\frac{\partial \rho}{ \partial p}=\frac{1}{c_s^2}=\left.\frac{\partial \rho}{\partial p}\right|_{p \neq p_z}+\Delta\rho\,\delta(p-p_z)~.
\label{eq:discontinuity}
\end{equation}
The $\Delta\rho\,\delta(p-p_z)$ term results in an extra contribution to the solution of $y(r)$, which can be written as
\begin{equation}
\Delta y=y(r_z^{+}) - y(r_z^{-}) = -\frac{4\pi r_z^{3} \, \Delta\rho}{m(r_z)+4\pi \,r_z^{3}\, p(r_z)}~~.
\label{eq:yjump}
\end{equation}

Here, $r_z$ denotes the position where $d\rho/dp$ becomes discontinuous, while $r_z^{\pm}$ represent two points infinitesimally close to $r_z$ from opposite directions. For an admixed quark star, $r_z$ corresponds to the radius of the $\OM$ core, denoted by $\R_{_\OM}$. Furthermore,
\begin{equation}
\Delta \rho = \rho\left(\R^{^-}_{_\OM}\right),
\end{equation}
represents the energy density immediately inside the region of quark matter. After acquiring the solution of the stellar structure equations, the corrected value of the perturbation function at the surface can be written as
\begin{equation}
y_{_R} = y(R) + \Delta y,
\end{equation}
where the inner boundary condition is taken as $y(0)=2$, irrespective of the presence of multiple fluids (see Refs.~\cite{Das_2022,Hinderer_2008}). Using the corrected value of $y_{_R}$, the quadrupolar electric-type tidal Love number $k_2$ can be calculated following Ref.~\cite{Hinderer_2008} as
\begin{eqnarray} \label{eq:k2} 
&& \hspace{-0.6cm} k_2 = \frac{8\C^5}{5} (1 - 2\C)^2 \left[ 2\C (y_{_R} - 1) - y_{_R} +2  \right] \nonumber\\
&& \hspace{-0.2cm} \times \Big[2C \big\{ + 3\C (5y_{_R} - 8) +6 -3y_{_R} \big\}\nonumber\\
&& \hspace{-0.2cm} + 4\C^3 \Big[ 13 + 2\C^2 (1 + y_{_R}) - 11y_{_R}+  \C(3y_{_R} - 2) \Big] \nonumber \\
&& +3(1 - 2\C)^2 \left[ 2 + 2\C (y_{_R}- y_{_R}- 1) \right] \ln(1 - 2\C) \Big]^{-1} \,.\end{eqnarray}
Finally, the dimensionless tidal deformability is obtained as
\begin{equation}
\label{eq:tidal}
\Lambda =\frac{2}{3} k_2\,\C^{-5},
\end{equation}
where $\C$ is the compactness of the corresponding stellar system.

\subsection{Universal Rotational-Tidal Relation}\label{Sec:IIId}
Finally, we investigate universal relations involving the dimensional moment of inertia and the tidal deformability of the two-fluid stellar system which is expressed in terms of the rotational-tidal $\left(\bar{\I}-\Lambda\right)$ response for two-fluid stellar configurations. This relation connects the moment of inertia, $\bar{\I} \coloneqq \I/\M^{3}$, with the tidal deformability, thereby providing a direct link between the microscopic descriptions of dense matter, encoded in the EoS, and the macroscopic observables of the corresponding stellar system. 

For single-fluid models, this relation (also known as $\I$-Love relation) is known to exhibit a remarkable insensitivity to the underlying EoS, yielding an almost universal curve that correlates the rotational and tidal responses of the stellar system. This universality can be exploited to place stringent constraints on the internal structure of NSs from observational data \cite{Kumar_2026}. In contrast, the presence of a second fluid introduces additional degrees of freedom that can modify both the rotational and tidal responses of the stellar configuration. In particular, the interaction between the two components alters the internal density and pressure profiles, which, in turn, can change both $\bar{\I}$ and $\Lambda$. It is therefore important to examine whether the universality of the $\bar{\I}$ and $\Lambda$ relation is preserved in two-fluid systems and, if so, to quantify the deviations from the single-fluid relation.
To quantify this relationship, we approximate the computed data with a single smooth curve through empirical fitting as
\begin{eqnarray}
    \log_{10}\bar{\I} = \sum^4_{n=0} \alpha_n \times\left(\log_{10} \Lambda\right)^n
\end{eqnarray}
 where $\alpha_n$ are the fitting coefficients determined from the numerical results. We also calculated the relative(or fractional) error given as
 \begin{eqnarray}
     \Delta = {|\bar{\I}-\bar{\I}_{\text{fit}}| \over \bar{\I}}
 \end{eqnarray}
The relative error deviation is less than $1\%$ \cite{Yagi_2013_365,Yagi_2013_023,Yagi_2017} for the single-fluid while for the two-fluid system, especially for the DM admixed NS, the relative error deviation can be greater than $1\%$ as shown by Cronin et al. \cite{Cronin_2023} or even $30\%$ as shown in the Ref. \cite{Kumar_2026}.

\section{Numerical Results and Discussions}\label{Sec:IV}
We provide the results and discussions of our analysis of two-fluid quark star systems using two OM EoS. We investigate the properties of both non-rotating or static and slowly rotating approximation stellar configurations, as well as the tidal deformability and universal relations of corresponding two-fluid stellar systems. Throughout this analysis, we kept the DM fraction aa $ 5\%$ and consider a range of bag constants, $\B_g = 60-75$ MeV/fm$^3$.

\subsection{Non-Rotating Case}
For the non-rotating case, we also provide with observational constraints from massive pulsar measurements, NICER observations, and GW detection events. The navy-colored and medium orchid colored bands provides the $1\sigma$ constraints of PSR J0348+0432 \cite{John_2013} and GW190814 \cite{Abbott_2020}. The red contour shows the NICER measurement data of PSR J0740+6620 \cite{Riley_2021}, while the $90\%$ CL of the PSR J0437-4715 \cite{Choudhury_2024} contour is presented in the purple color. The blue color contour represents the PSR J0030+0451 \cite{Riley_2019}. The LIGO-Virgo detections of GW events such as GW190425 \cite{Abbott_2020_L3} and GW170817 \cite {Abbott_2018_16} binary NS mergers are shown, respectively, in green and gold. In addition, we included the contour of $68\%$ CL for HESS J1731-347 \cite{Doroshenko_2022} with black color. 
\begin{figure*}[ht]
    \centering
    \begin{minipage}{0.48\textwidth}
        \centering
        \includegraphics[width=\linewidth,height=7cm]{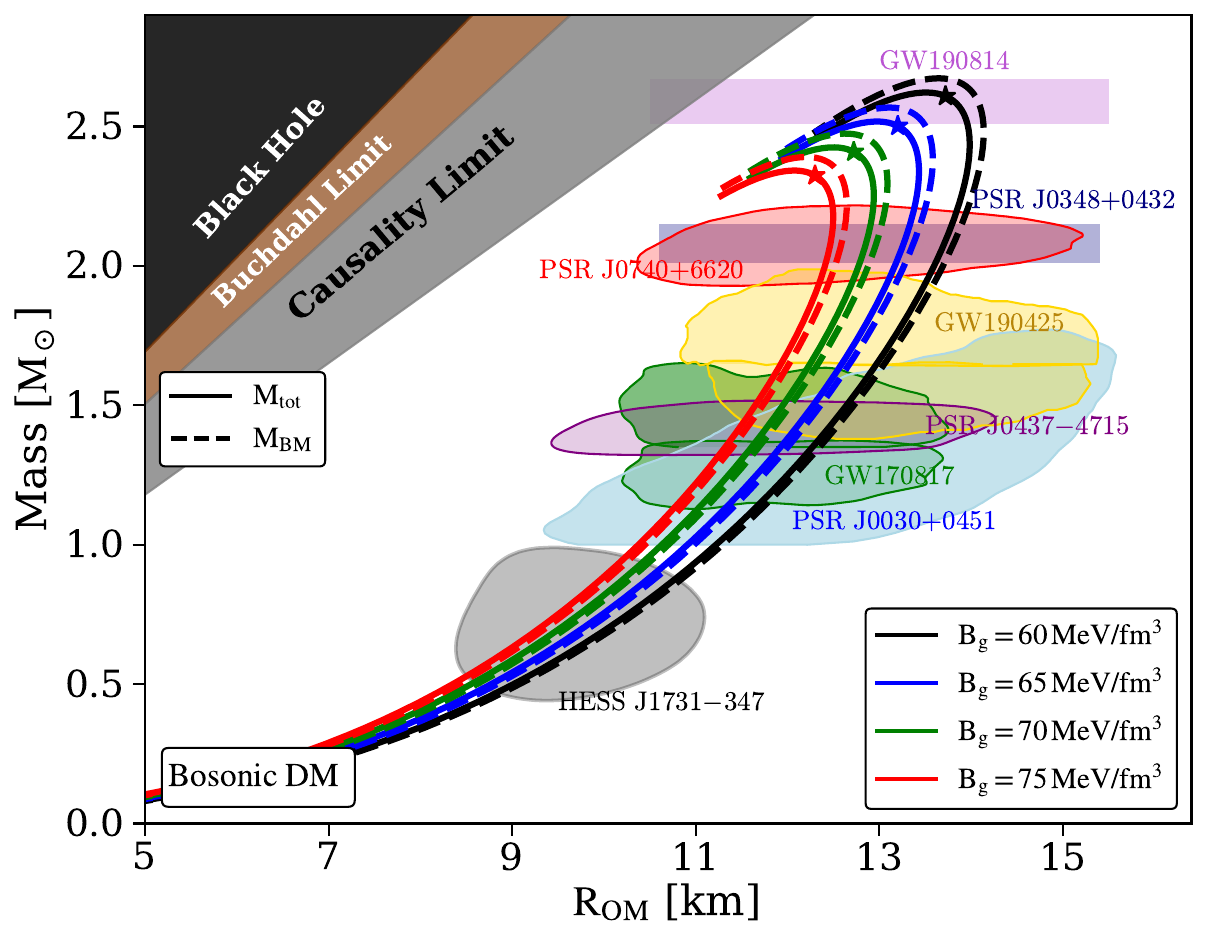}
    \end{minipage}
    \hfill
    \begin{minipage}{0.48\textwidth}
        \centering
        \includegraphics[width=\linewidth,height=7cm]{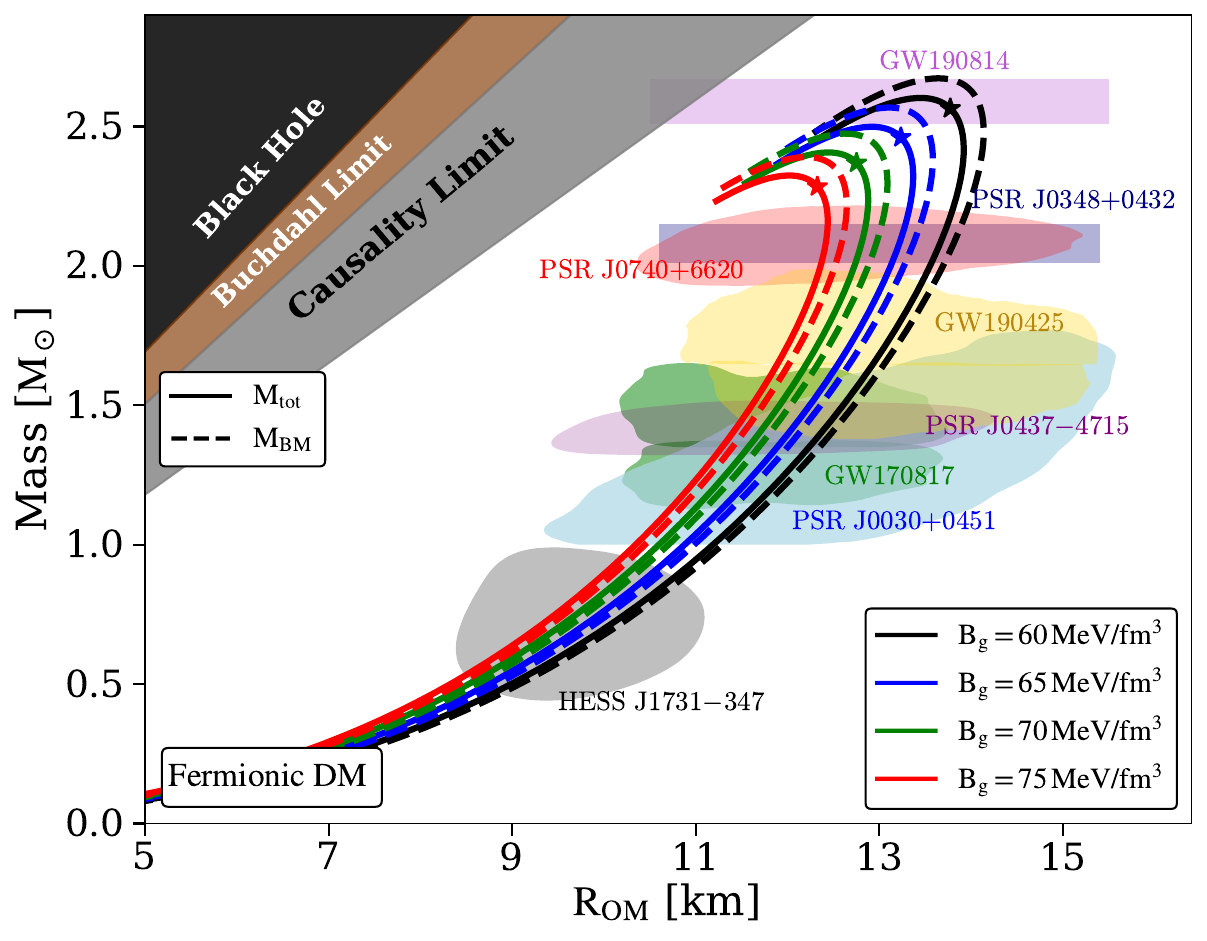}
    \end{minipage}
    \caption{\justifying $\M-\R$ relations for quark star described by BM EoS admixed bosonic DM (left) and fermionic DM (right) for bag constants $\B_g=60,\ 65,\ 70,$ and $75~\mathrm{MeV/fm^3}$ and DM fraction,$f_\x=5\%$. The shaded regions indicate the corresponding observational constraints.}
    \label{fig:MRBM}
\end{figure*}

\begin{table*}[ht]
    \caption{
Maximum-mass configurations for quark stars described by BM admixed with bosonic and fermionic DM for the different bag constants, asumming $f_\x =5\%$.
}\label{tab:MRBM}
\centering
\begin{tabular}{c|cc|cc|cc}
\hline
& \multicolumn{2}{c|}{BM EoS} &
\multicolumn{2}{c|}{Bosonic DM} &
\multicolumn{2}{c}{Fermionic DM}\\
\cline{2-7}
$\B_g$ (MeV/fm$^{3}$)
& $\M_{\BM}$ $(\Ms)$
& $\R_{\OM}$ (km)
& $\M_{\tot}$ $(\Ms)$
& $\R_{\OM}$ (km)
& $\M_{\tot}$ $(\Ms)$
& $\R_{\OM}$ (km)\\
\hline
60  & 2.673 & 13.428 & 2.622 & 13.507 & 2.602 & 13.417 \\
65  & 2.568 & 13.117 & 2.518 & 12.967 & 2.499 & 12.873 \\
70  & 2.475 & 12.627 & 2.425 & 12.497 & 2.407 & 12.472 \\
75  & 2.391 & 12.199 & 2.342 & 12.070 & 2.324 & 12.024 \\
\hline
\end{tabular}
\end{table*}

The $\M-\R$ relationships for BM quark stars admixed with bosonic DM and fermionic DM are diplayed in Fig.~\ref{fig:MRBM}. These results are presented for various bag constant values $\B_g$, while keeping the DM fraction fixed at $f_\x = 5\%$. The solid curves represent the total gravitational mass configurations, whereas the dashed curves correspond to the ordinary matter component. In both scenarios, an increase in the bag constant results in a shift of the mass and radius sequences toward lower maximum masses and reduced stellar radii. Typically, stars with bosonic DM can support larger maximum masses and radii, while those with fermionic DM tend to be more compact. This distinction arises from the unique EoS and pressure support mechanisms associated with bosonic and fermionic DM. The maximum masses and their respective radii are presented in the Table \ref{tab:MRBM}.

\begin{figure*}[ht]
\centering
    \begin{minipage}{0.48\textwidth}
        \centering
        \includegraphics[width=\linewidth,height=7cm]{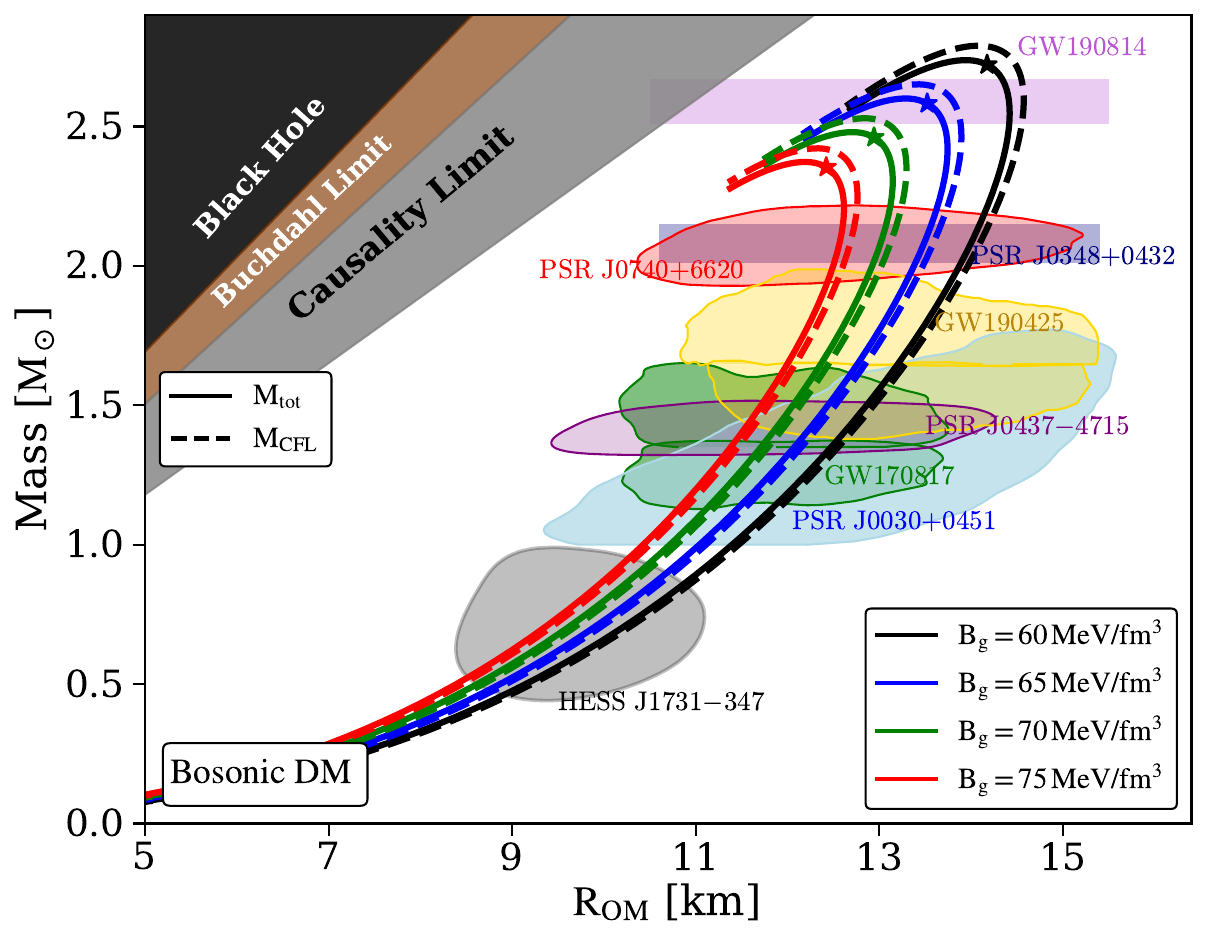}
    \end{minipage}
    \hfill
    \begin{minipage}{0.48\textwidth}
        \centering
        \includegraphics[width=\linewidth,height=7cm]{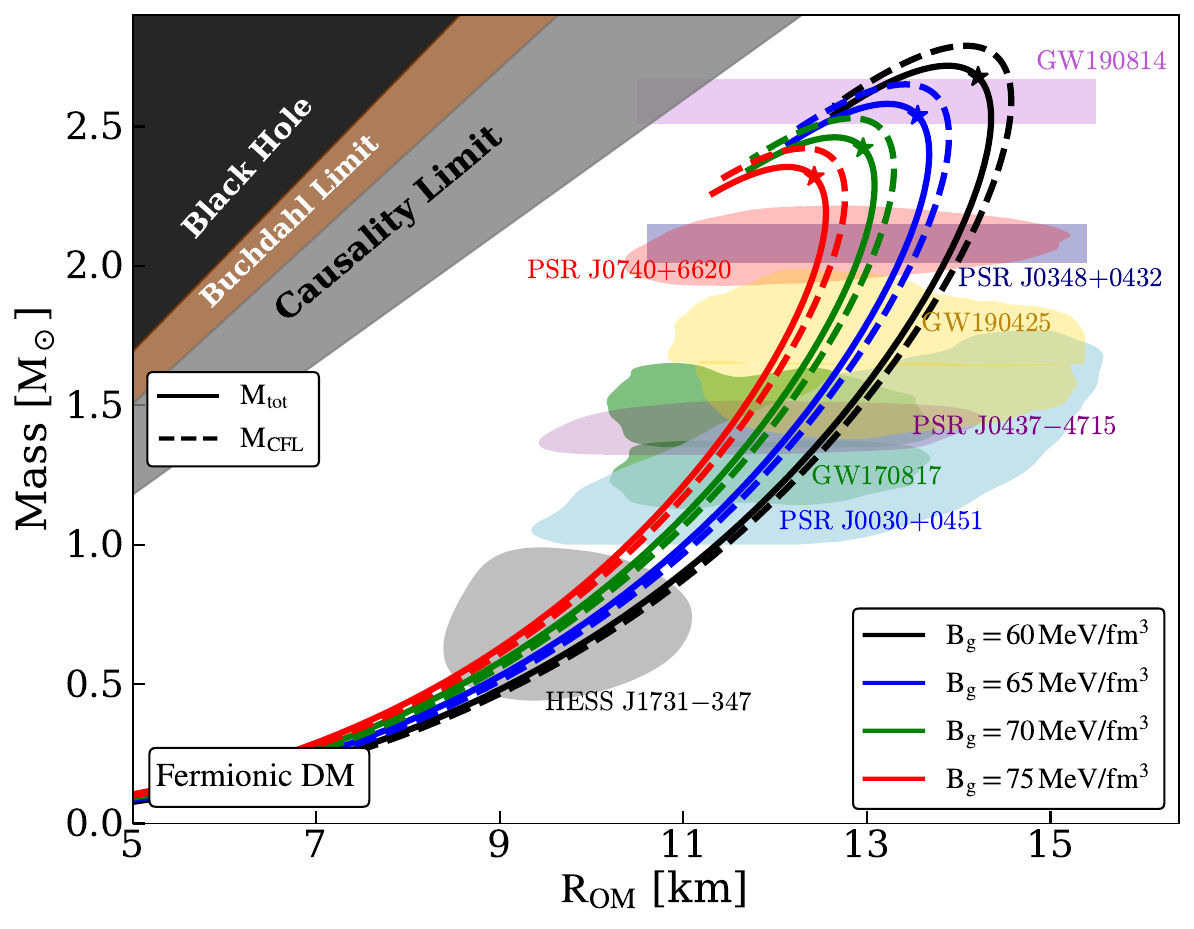}
    \end{minipage}
    \caption{\justifying $\M-\R$ relations for quark star described by CFL EoS admixed bosonic DM (left) and fermionic DM (right) for the same bag constants and DM fraction,$f_\x=5\%$. The shaded regions indicate the corresponding observational constraints.}
    \label{fig:MRCFL}
\end{figure*}

\begin{table*}[ht]
\caption{
Maximum-mass configurations for quark stars described by CFL EoS admixed with bosonic and fermionic DM for the different bag constants, asumming $f_\x =5\%$.
}
\label{tab:CFLMR}
\centering
\begin{tabular}{c|cc|cc|cc}
\hline
& \multicolumn{2}{c|}{BM EoS} & \multicolumn{2}{c|}{Bosonic DM}
& \multicolumn{2}{c}{Fermionic DM} \\
\cline{2-7}
$\B_g$ (MeV/fm$^{3}$) & $\M_{_\BM}$ $(\Ms)$ & $\R_{_\OM}$ (km)
& $\M_{\tot}$ $(\Ms)$ & $\R_{_\OM}$ (km) & $\M_{\tot}$ $(\Ms)$
& $\R_{_\OM}$ (km)\\
\hline
60  & 2.790 & 14.047 & 2.738 & 13.931 & 2.718 & 13.913 \\
65  & 2.651 & 13.410 & 2.601 & 13.294 & 2.581 & 13.251 \\
70  & 2.530 & 12.840 & 2.480 & 12.725 & 2.461 & 12.661 \\
75  & 2.422 & 12.329 & 2.373 & 12.215 & 2.355 & 12.130 \\
\hline
\end{tabular}
\end{table*}

Fig.~\ref{fig:MRCFL} illustrates the $\M-\R$ plots for CFL EoS admixed with bosonic DM and fermionic DM. For both types of DM, an increase in the bag constant decreases the maximum stellar mass and tends the equilibrium configurations toward smaller radii, indicating a gradual softening of the CFL EoS. A comparison of the two panels shows that configurations with bosonic DM generally support slightly higher maximum masses and radii than those with fermionic DM. Alternatively, the presence of fermionic DM tends to result in denser stellar configurations due to its relatively softer effective EoS.

The deviation the total gravitational mass curves from the respective OM branches becomes increasingly evident near the maximum-mass configurations. This illustrates that even a small fraction of DM can significantly affect the overall equilibrium structure of CFL compact stars. Nevertheless, the selected DM fraction of $f_\x = 5\% $ preserves the fundamental characteristics of the underlying strange star configurations. Table \ref{tab:CFLMR} displays the maximum masses and respective radii for OM, bosonic DM, and fermionic DM.

When compared to the BM EoS, the CFL EoS tends to result in more compact stellar structures due to pairing effects among quarks. As a result, the $\M-\R$ relationships show larger stellar radii and altered maximum mass configurations. The inclusion of DM further modifies the equilibrium structure of CFL stars, leading to additional deviations from the OM branch. It is also noted that both BM and CFL models with lower bag constants meet the observational criterion of supporting compact stellars beyond $2.5\,\M_\odot$, which is relevant for constraining GW190814, while still aligning with the recent constraints from GW190425, PSR J0740$+$6620,PSR J0348$+$0432 and GW170817.

Further, we briefly analyzed the stability of the stellar system. For the one-fluid system, the stability of a stellar structure can be assessed by examining radial perturbations. To achieve this, a harmonic perturbation of the form $\xi(r)\,e^{-i w t}$ is introduced into both the metric and fluid variables, where $\xi(r)$ represents the radial displacement and $w$ denotes the eigenfrequency of the oscillation mode. By linearizing the field equations around the equilibrium state, we obtain a radial perturbation equation, which can be framed as a Sturm-Liouville eigenvalue problem for $w^{2}$. The solutions to this differential equation reveal the stability of the stellar configuration. Specifically, when $w^{2} > 0$, the configurations correspond to stable oscillatory modes, implying that the star maintains dynamic stability against small radial perturbations. Conversely, configurations with $w^{2} < 0$ indicate unstable modes, in which perturbations grow exponentially over time, potentially leading to the gravitational collapse of the stellar structure. For a single-fluid, the position of the last stable configuration, indicated by $(w = 0)$, aligns with the maximum mass in the $\M-\R$ diagram. This can be defined by the condition $\partial M/\partial \rho^c = 0$, where $\rho^c$ represents the central energy density.

In the context of the two-fluid formalism, the generalized stability criterion indicates that, at the point of instability, the quantities of ordinary matter particles $\N_\OM$ and DM particles $\N_\DM$ remain unchanged despite fluctuations in the central energy densities $\rho_\OM^c$ and $\rho_{_\DM}^c$~\cite{Hippert_2023,Marzola_2025}. The $\N_{_\OM}$ and $\N_{_\DM}$, respectively, can be calculated from the given differential equations by solving simultaneously with the fluid TOV equation
\begin{eqnarray}
   {d \N_{_\OM}\over dr} &=& {4\pi r^2 \rho_{_\OM} \over \sqrt{1- 2m_\tot/r}} \\
   {d\N_{_\DM} \over dr} &=& {4\pi r^2 \rho_{_\DM} \over \sqrt{1- 2m_\tot/r}}~~.
\end{eqnarray}  
The stability analysis of the two-fluid system is conducted by examining the diagonal elements of the matrix  $\partial \N_x / \partial \rho_x^c$ with $(x,y = \OM,\DM)$ from the system of equations as in Refs.~\cite{Barbat_2024,Hippert_2023,Biesdorf_2025}
\begin{equation}
\begin{pmatrix}
\delta \N_{_\OM} \\
\delta \N_{_\DM}
\end{pmatrix}
=
\begin{pmatrix}
\dfrac{\partial \N_{_\OM}}{\partial \rho^{c}_{_\OM}} ~&~
\dfrac{\partial \N_{_\OM}}{\partial \rho^{c}_{_\DM}}
\\[2.2ex]
\dfrac{\partial \N_{_\DM}}{\partial \rho^{c}_{_\OM}} ~&~
\dfrac{\partial \N_{_\DM}}{\partial \rho^{c}_{_\DM}}
\end{pmatrix}
\begin{pmatrix}
\delta \rho^{c}_{_\OM} \\
\delta \rho^{c}_{_\DM}
\end{pmatrix}
=0~~ ,
\end{equation}
and the configuration is stable only when eigenvalues are  positive i.e.,
\begin{eqnarray}
\label{eq:evalues}
    \kappa_1 >0 \quad \mbox{and} \quad \kappa_2 >0
\end{eqnarray}
 which are the generalized conditions of a multi-fluid system. As based on the various studies \cite{Marzola_2025,Leung_2012,Leung_2022}, we consider a fixed DM fraction of $f_\x = 5\%$ and, for each given value of $\B_g$, we pinpoint the maximum mass configuration that corresponds to the last stable star by employing the condition given in Eq.~\eqref{eq:evalues}. These configurations are indicated by star symbols in the $\M-\R$ plots depicted in Figs.~\ref{fig:MRBM} and \ref{fig:MRCFL}. It is noted that with a fixed $f_\x$, the last stable mass closely aligns with the maximum mass configuration for each equilibrium sequence, indicating that most stellar configurations before the maximum mass are dynamically stable for the corresponding system. 

For bosonic DM, the difference between the maximum mass and the last stable mass is nearly negligible. This indicates that with the small values of $f_\x$, bosonic DM does not significantly alter the stability region of compact star configurations. As a result, the onset of instability occurs almost exactly at the maximum mass configuration, similar to the behavior observed in the standard single-fluid case.

In contrast, for fermionic DM, the last stable mass is slightly less than the corresponding maximum mass. This suggests that instability begins just before the equilibrium sequence reaches its absolute maximum mass. Although the deviation is small, it indicates that fermionic DM has a somewhat stronger effect on the stability of the system. Nevertheless, the overall structure of the equilibrium sequences remains largely intact. Moreover, in the case where only the OM EoS is considered, the last stable mass precisely coincides with the maximum mass for each curve. This is consistent with the conventional stability criterion for a single-fluid stellar configuration. 

\subsection{Slowly Rotating Case}

For slowly rotating stellar configurations, the same EoS and model parameters used in the static analysis are applied. In this work, we consider the ratio of the total angular velocity of each fluid $f_{_\Omega} = \Omega_{_\OM} / \Omega_{_\DM}$ as unity and calculate the moment of inertia as a function of the gravitational mass ($\I-\M$) relation.

\begin{figure}[!htp]
    \centering   
   \includegraphics[width=\linewidth,height=7cm]{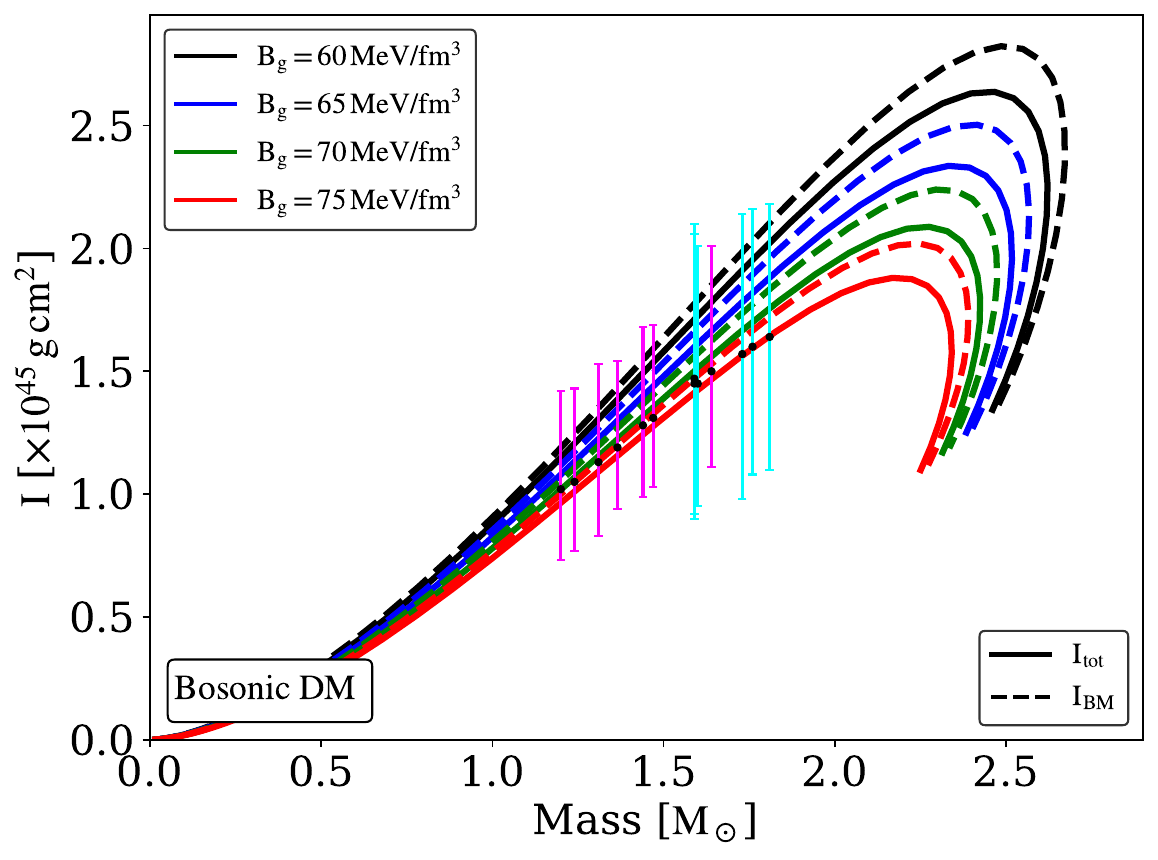} 
    \vspace{0.3cm}
    \includegraphics[width=\linewidth,height=7cm]{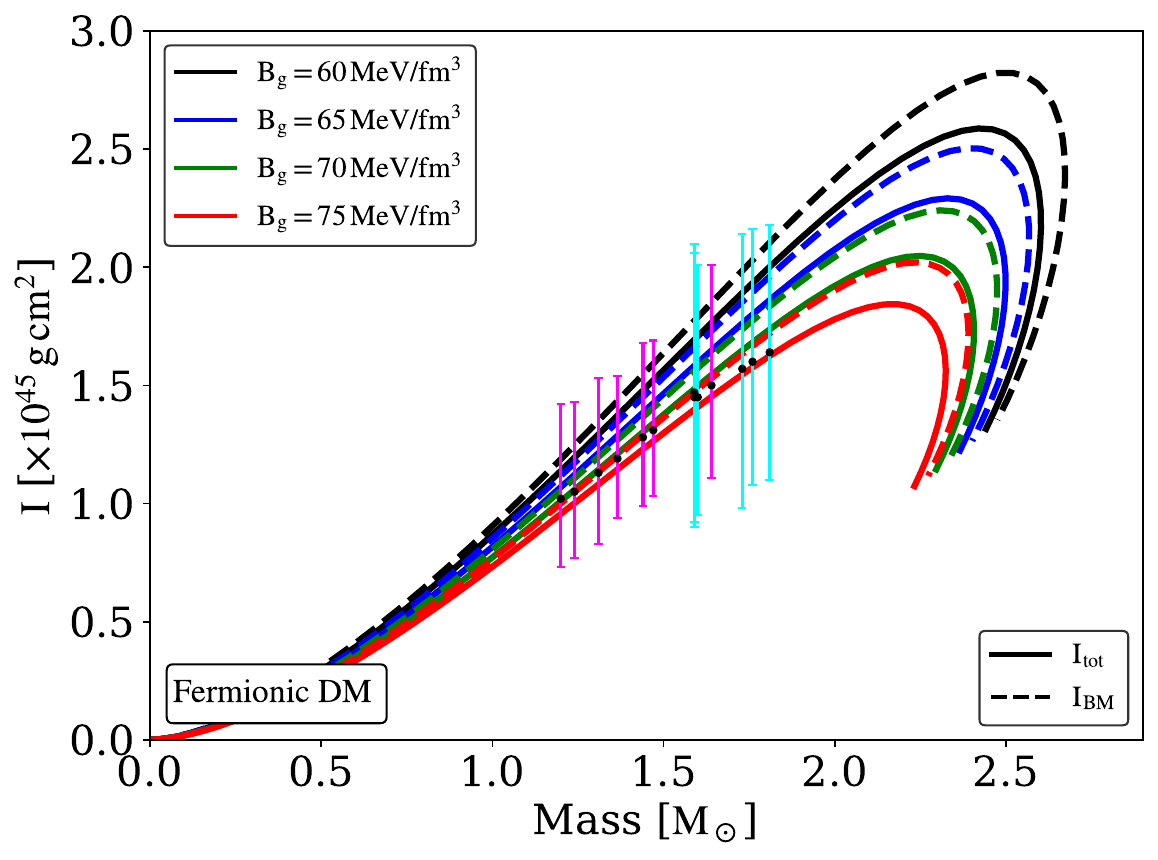}
    \caption{\justifying $\I-\M$ relations for quark star described by BM EoS admixed bosonic DM (upper-panel) and fermionic DM (lower-panel) for same configurations. The error bars with magenta and cyan colors show the inferred ranges of the moment of inertia from millisecond pulsars and low-mass X-ray binaries, respectively, and indicate the corresponding observational constraints.}
    \label{fig:IMBM}
\end{figure}

Fig.~\ref{fig:IMBM} $\I-\M$ relation for a slowly rotating stellar system for BM admixed with bosonic and fermionic DM for different values of the bag constant. Generally, moment of inertia increases with stellar mass, peaking near maximum mass configurations, as more massive stars have greater rotational inertia. Lower bag constant values correspond to larger moment of inertia, while higher $\B_g$ values reduce them due to a softer EoS, resulting in more compact stars with smaller radii. The OM component consistently shows a higher moment of inertia than the total (i.e., $\I_\BM>\I_\tot$), underscoring the role of DM. For both DM, deviations of inclusion of DM EoS are clearly observed, resulting in substantial modifications to the internal mass distribution and rotational characteristics.

\begin{figure}[!htp]
    \centering
   \includegraphics[width=\linewidth,height=7cm]{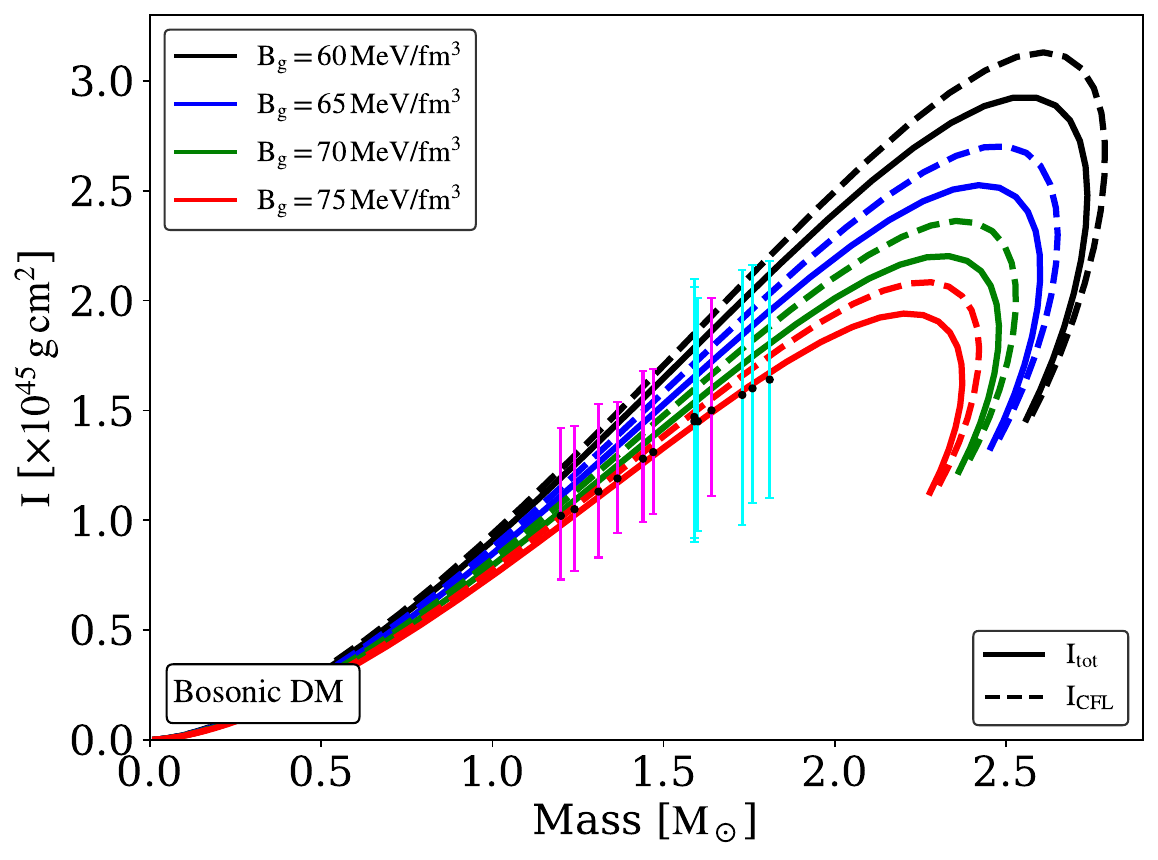} 
    \vspace{0.3cm}
    \includegraphics[width=\linewidth,height=7cm]{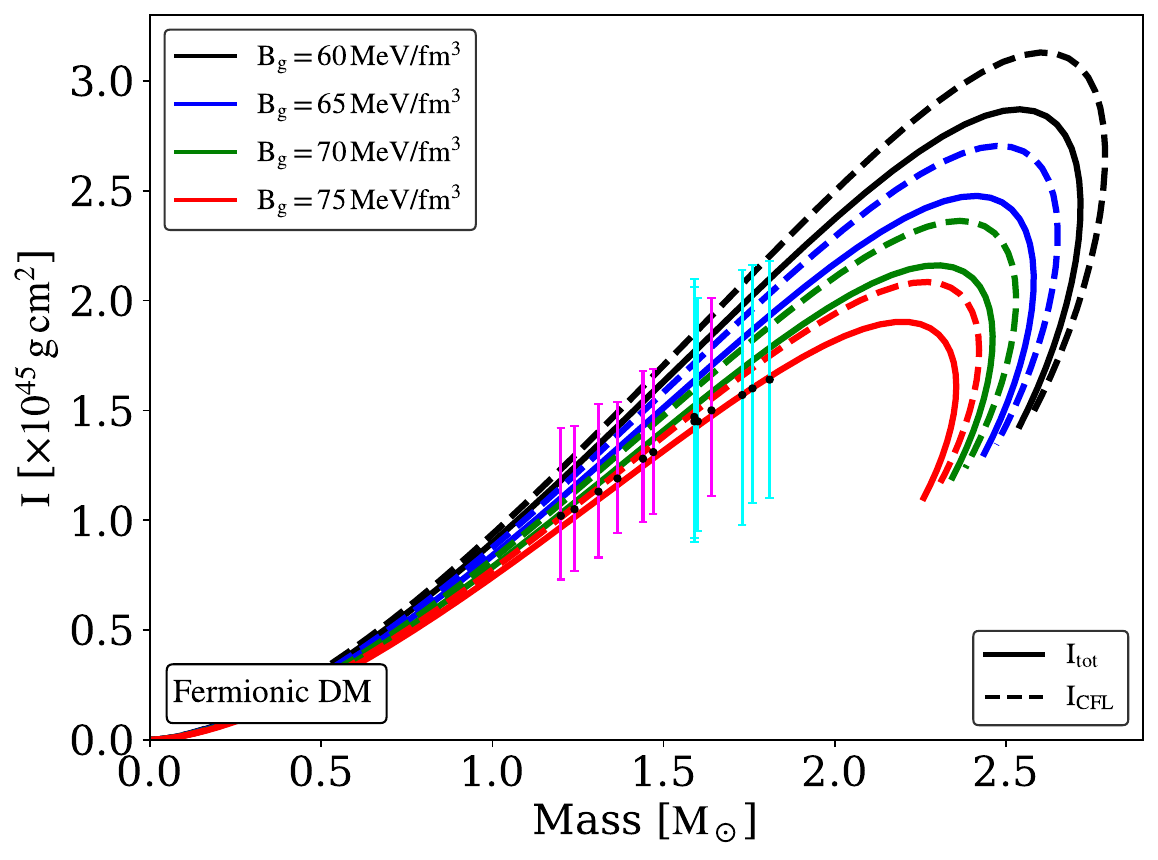}
    \caption{\justifying $\I-\M$ relations for quark star described by CFL EoS admixed bosonic DM (upper-panel) and fermionic DM (lower-panel) for same configurations. The error bars with magenta and cyan colors show the inferred ranges of the moment of inertia from millisecond pulsars and low-mass X-ray binaries, respectively, and indicate the corresponding observational constraints.}
    
    \label{fig:IMCFL}
\end{figure}

Fig.~\ref{fig:IMCFL} shows the $\I-\M$ relations for the CFL stellar system admixed with bosonic and fermionic DM for different values of the bag constant. The CFL EoS typically yields larger maximum masses and moments of inertia than the BM. This indicates that the CFL phase supports stiffer stellar structures, enhancing their rotational stability. Similar to the BM phase, the OM component remains higher than the total moment of inertia curves, highlighting the significant contribution of DM to the overall rotational properties of the stellar system.

When comparing bosonic and fermionic DM, the deviations in total and OM only moments of inertia are clearly observed for the both DM EoS. Overall, the CFL configurations not only support greater maximum masses but also exhibit larger moments of inertia than BM configurations, due to their stiffer nature and enhanced rotational support.

Finally, the CFL configuration typically supports larger maximum masses and greater moments of inertia than the corresponding BM configuration, given the same values for the bag constant and DM fraction. This behaviour suggests that the CFL EoS yields stiffer stellar configurations, thereby enhancing rotational support and increasing moment of inertia. Further, Figs.~\ref{fig:IMBM} and \ref{fig:IMCFL} show that current models of DM admixed with quark stars are consistent with the inferred ranges of millisecond pulsar moments of inertia (magenta color) and of NS low-mass X-ray binaries (cyan color) provided in the Ref.~ \cite{Kumar_2019}. The predicted $\I-\M$ relations for both BM and CFL EoS align with millisecond pulsar measurements, indicating that the DM inclusion does not significantly violate current observational bounds.

Additionally, the larger moment of inertia associated with the CFL EoS is consistent with current astrophysical constraints. As the moment of inertia is sensitive to internal structure and stellar radius, future high-precision measurements of pulsar properties could provide valuable constraints on the EoS and the presence of DM in compact quark stars.

\begin{figure}[!htp]
    \centering   
   \includegraphics[width=\linewidth,height=7cm]{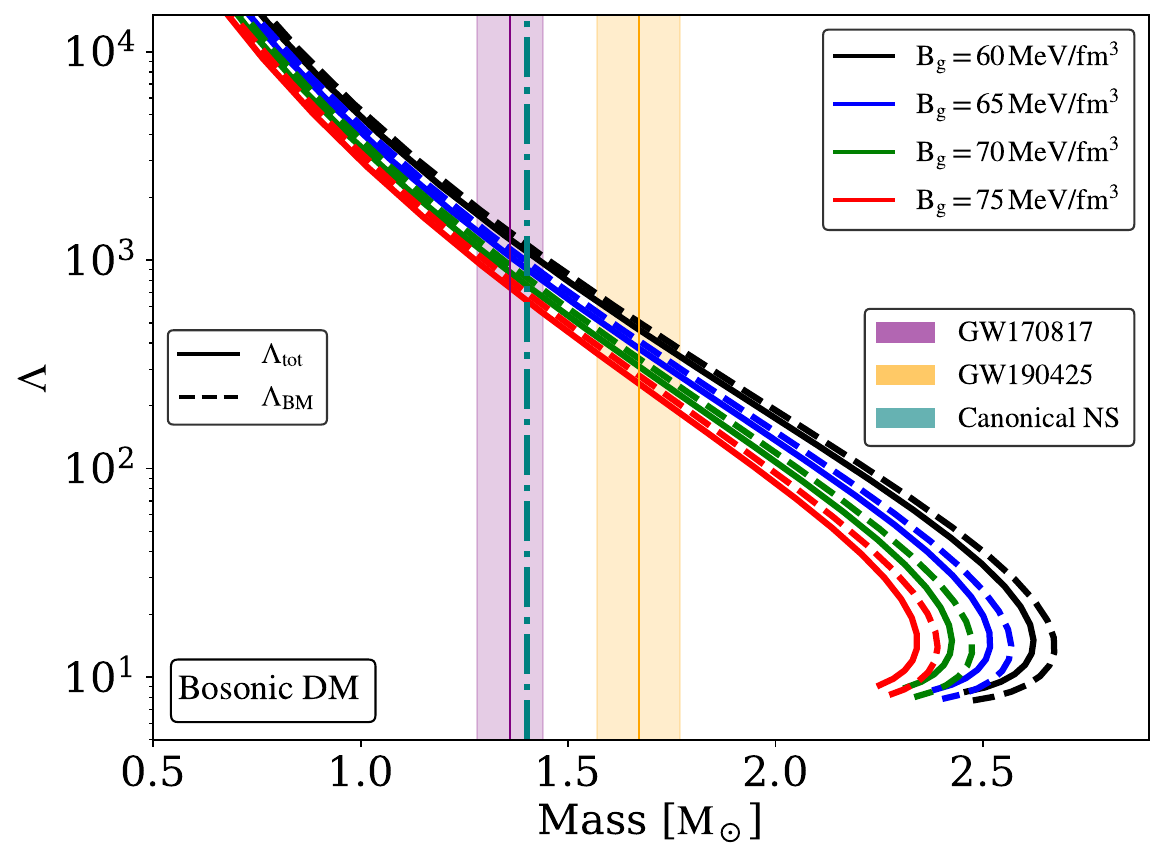}
    \vspace{0.3cm}
    \includegraphics[width=\linewidth,height=7cm]{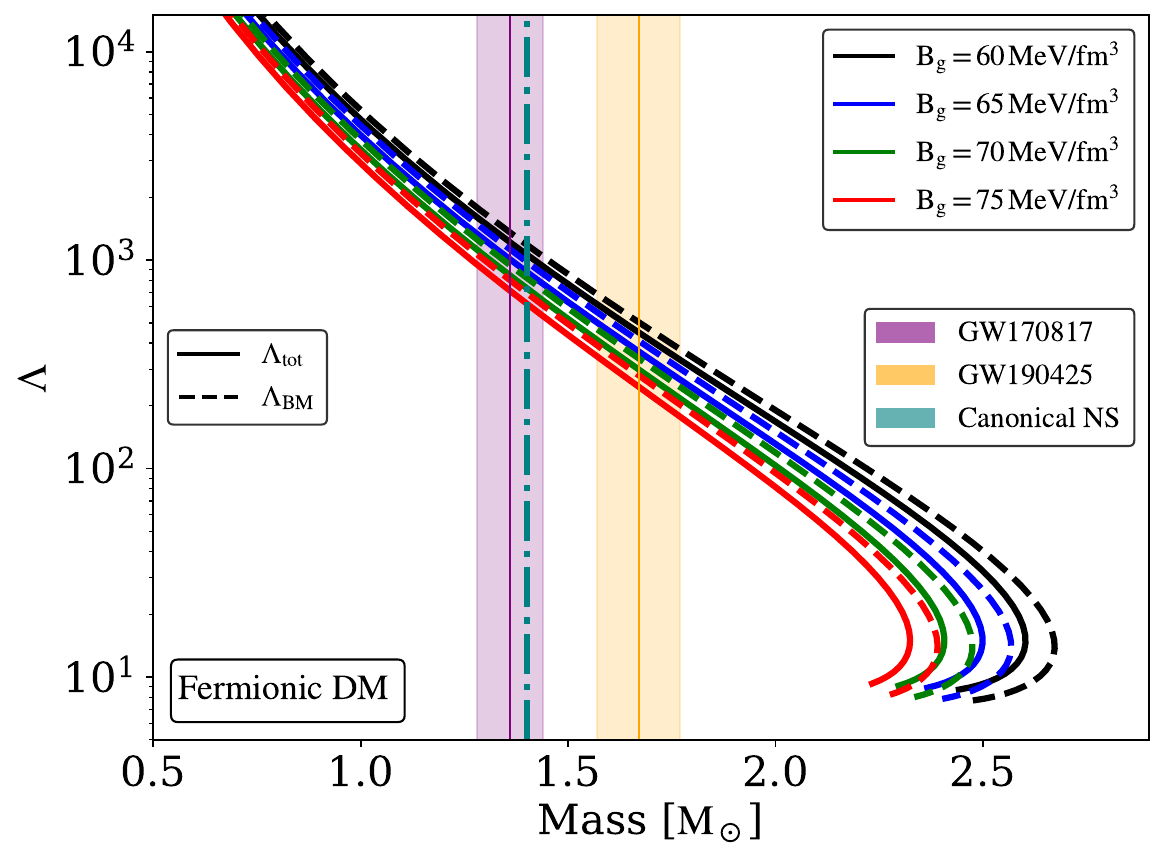}
    \caption{\justifying $\Lambda-\M$ relations for quark star described by BM EoS admixed bosonic DM (upper-panel) and fermionic DM (lower-panel) for same configurations. The shaded vertical bands indicate the observational constraints associated with GW170817, GW190425, and canonical NS masses.} 
    \label{fig:TidalBM}
\end{figure}

\begin{figure}[!htp]
    \centering
   \includegraphics[width=\linewidth,height=7cm]{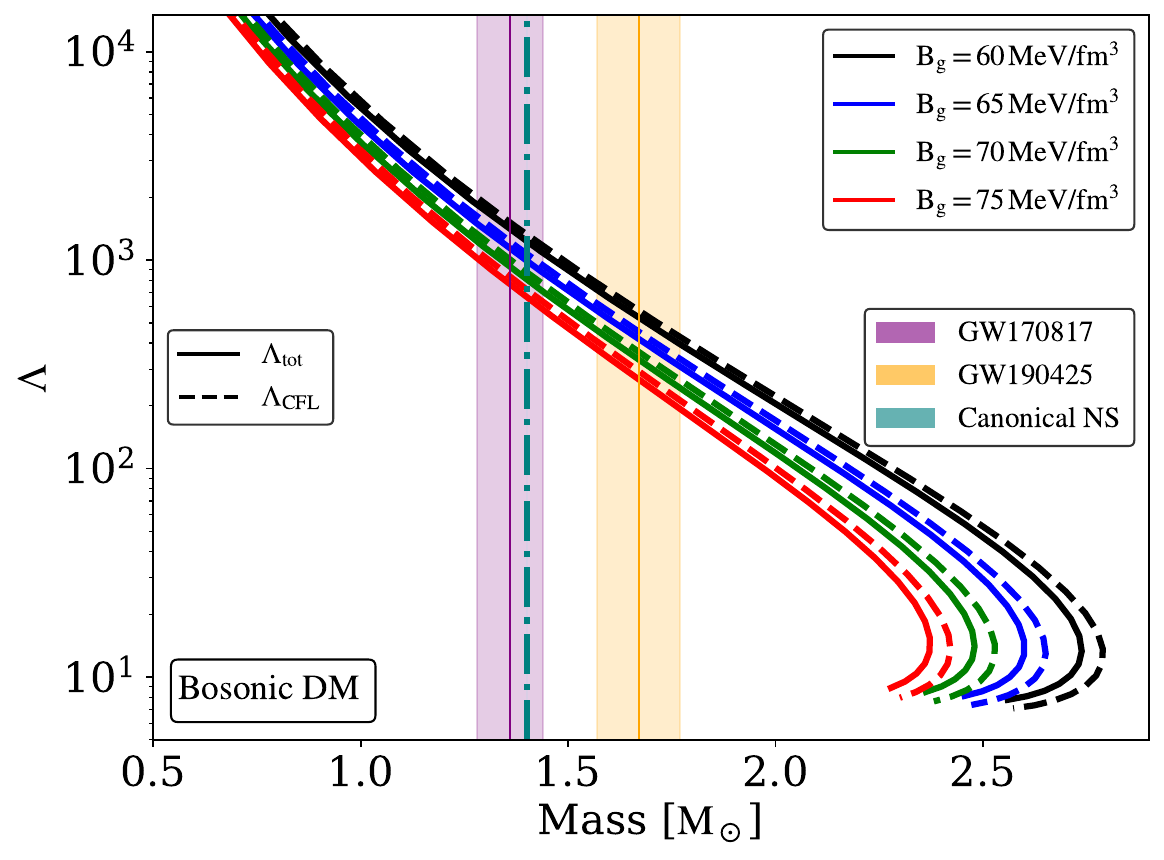}
    \vspace{0.3cm}
    \includegraphics[width=\linewidth,height=7cm]{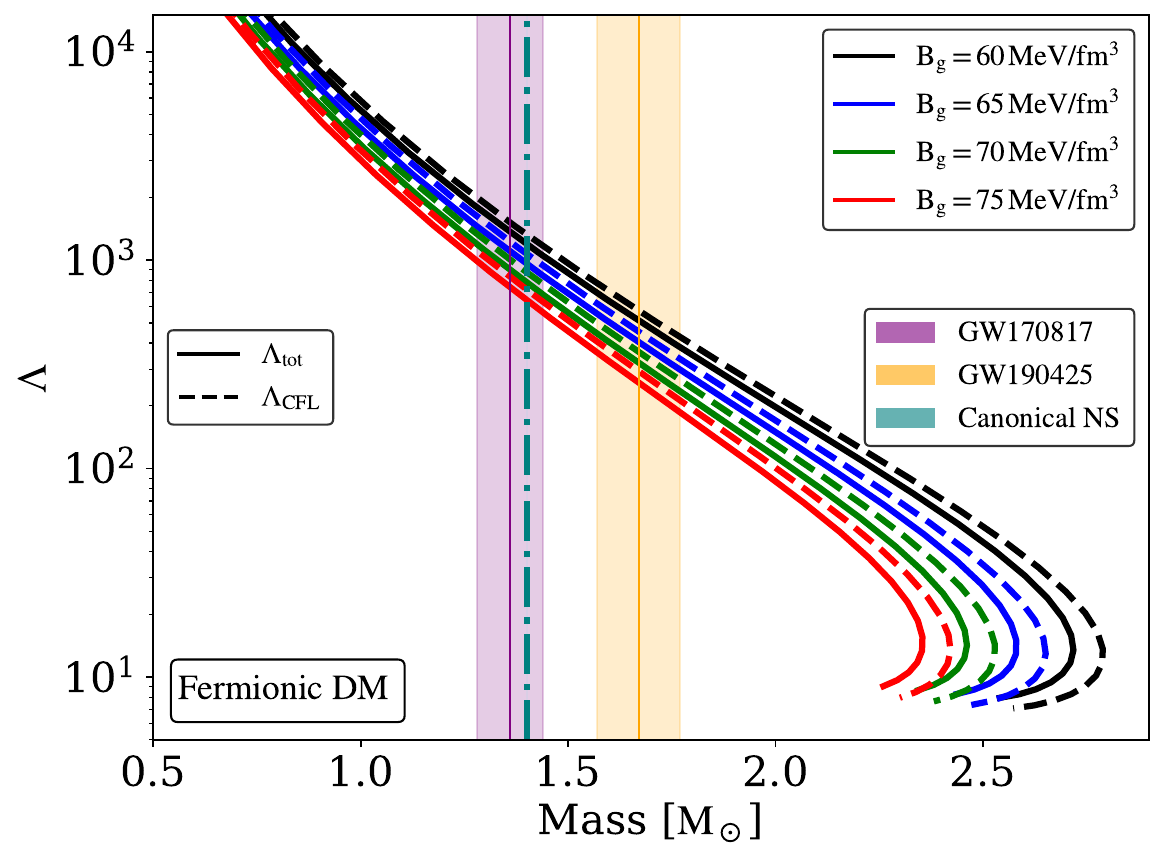}
    \caption{$\Lambda-\M$ relations for quark star described by CFL EoS admixed bosonic DM (upper-panel) and fermionic DM (lower-panel) for same configurations. The shaded vertical bands indicate the observational constraints associated with GW170817, GW190425, and canonical NS masses.}
    \label{fig:TidalCFL}
\end{figure}

\subsection{Tidal Deformability}

We also calculated the dimensionless tidal deformability, $\Lambda$, as a function of the gravitational mass $\M$. Using the same configuration, we examined how consistent our models are with the tidal deformability constraints from canonical NS, as well as from GWs events. The purple and orange shaded regions are  $1-\sigma$ \cite{Chatziioannou_2020} constraint on the binary system of GW events GW170817 \cite{Abbott_2017_18} and GW190425 \cite{Abbott_2020_L3} respectively, while teal color dotdashed line represents the canonical NS ($\Lambda_{1.4}$) \cite{Hinderer_2010}. 

Fig.~\ref{fig:TidalBM} illustrates the $\Lambda-\M$ relations for the BM admixed with bosonic and fermionic $\DM$ across different values of the bag constant. In all scenarios, the tidal deformability decreases monotonically as the gravitational mass increases. Furthermore, increasing the bag constant systematically reduces the tidal deformability of the stellar system. For both the bosonic and fermionic DM EoS, the total tidal deformability, $\Lambda_\tot$, remains consistently lower than the tidal deformability of the BM, $\Lambda_\BM$, for all values of $\B_g$ considered. This shows that including a $\DM$ fraction of $f_\x = 5\%$ slightly suppresses the tidal response of the star. However, the difference between $\Lambda_\tot$ and $\Lambda_\BM$ is relatively small, indicating that this $5\%$ contribution has only a weak effect on the star's tidal response. The corresponding values of tidal deformability for the canonical NS, GW170817, and GW190425 are detailed in Table~\ref{tab:MITLambda}.

%%%%%%%%%%%%%%%%%%%%%%%%%%%%%%%%%%%%%%%%%%%%

\begin{table*}[!htp]
\centering
\caption{ Dimensionless tidal deformability $\Lambda$ for quark stars described by BM EoS admixed with bosonic and fermionic DM for the different bag constants, asumming $f_\x =5\%$. The values of $\Lambda$ are shown for the canonical mass configuration and for the masses of GW170817 and GW190425.}
\small
\setlength{\tabcolsep}{4pt}
\resizebox{\textwidth}{!}{
\begin{tabular}{c|ccc|ccc|ccc}
\hline

&
\multicolumn{3}{c|}{BM EoS}
&
\multicolumn{3}{c|}{Bosonic DM}
&
\multicolumn{3}{c}{Fermionic DM}
\\

\cline{2-10}

$\B_g$ (MeV/fm$^{3}$)

& Canonical NS
& GW170817
& GW190425

& Canonical NS
& GW170817
& GW190425

& Canonical NS
& GW170817
& GW190425
\\

\hline

60
& $ 1230$ & $1369$ & $506$
& $1134$ & $1269$ & $467$
& $1076$ & $1224$ & $454$
\\

65
& $993$ & $1139$ & $413$
& $925$ & $1053$ & $4379$
& $884$  & $1019$ & $365$
\\

70
& $826$ & $957$ & $340$
& $758$ & $899$ & $314$
& $807$ & $930$ & $331$
\\

75
& $703$ & $803$ & $282$
& $657$ & $762$ & $261$
& $625$ & $716$ & $247$
\\

\hline
\end{tabular}
}

\label{tab:MITLambda}
\end{table*}

\begin{table*}[!htp]
\centering
\caption{ Dimensionless tidal deformability $\Lambda$ for quark stars described by CFL EoS admixed with bosonic and fermionic DM for the different bag constants, asumming $f_\x =5\%$. The values of $\Lambda$ are shown for the canonical mass configuration and for the masses of GW170817 and GW190425.
}
\small
\setlength{\tabcolsep}{4pt}

\resizebox{\textwidth}{!}{
\begin{tabular}{c|ccc|ccc|ccc}
\hline

&
\multicolumn{3}{c|}{BM EoS}
&
\multicolumn{3}{c|}{Bosonic DM}
&
\multicolumn{3}{c}{Fermionic DM}
\\

\cline{2-10}

$\B_g$ (MeV/fm$^{3}$)

& Canonical NS
& GW170817
& GW190425

& Canonical NS
& GW170817
& GW190425

& Canonical NS
& GW170817
& GW190425
\\

\hline

60
& $1355$ & $1536$ & $588$
& $1270$ & $1467$ & $547$
& $1229$ & $1422$ & $529$
\\

65
& $1089$ & $1250$ & $436$
& $1022$ & $1181$ & $431$
& $989$  & $1144$ & $416$
\\

70
& $880$ & $1034$ & $370$
& $835$ & $961$ & $342$
& $852$ & $958$ & $343$
\\

75
& $732$ & $862$ & $299$
& $688$ & $786$ & $275$
& $695$ & $807$ & $276$
\\

\hline
\end{tabular}
}

\label{tab:CFLLambda}
\end{table*}

\begin{table*} [!htp]
    \centering
    \setlength{\tabcolsep}{4pt}
  \caption{\label{tab:ILoveCoeffs}
Fitting coefficients for the universal $\I$--Love relations,
$\log_{10}\bar{\I}=\sum_{n=0}^{4}\alpha_n \times
(\log_{10}\Lambda)^n$, for the stellar configurations obtained by admixing $f_\x=5\%$  with the BM and CFL EoS with bosonic and fermionic DM and for BM and CFL EoS only with various values of $\B_g$}
    {
    \begin{tabular}{l|r|r|r|r|r}
    \hline
      Model      & $\alpha_0$   & $\alpha_1$   & $\alpha_2$    & $\alpha_3$    & $\alpha_4$    \\
      \hline
      BM+Bosonic DM & $-0.111586$ &  $0.610081$  &  $-0.137899$ &  $0.027376$ &  $-0.001804$ \\
      BM+Fermionic DM  & $-0.110464$  & $0.611972$   & $-0.139556$   & $0.027777$    & $-0.001834$   \\
      CFL+Bosonic DM & $-0.017084$ &  $0.451738$  &  $-0.051819$&  $0.009075$ &  $-0.000488$ \\
      CFL+Fermionic DM   & $-0.113939 $ & $0.612363$  & $-0.138669$  & $0.027500$  & $-0.000486$ \\
      \hline
    \end{tabular}}

\end{table*}

%%%%%%%%%%%%%%%%%%%%%%%%%%%%%%%%%%%%%%%%%%%%%%%%%%%%%

Furthermore, Fig.~\ref{fig:TidalCFL} presents the $\Lambda-\M$ relations for the CFL EoS admixed with fermionic and bosonic DM EoS. As in the BM case, tidal deformability again decreases monotonically with the increase in the gravitational mass, and higher bag constants lead to systematically smaller tidal deformabilities. The total tidal deformability, $\Lambda_\tot$, remains slightly lower than the CFL EoS contribution, $\Lambda_{_\CFL}$, which suggests that the $\DM$ component also suppresses the tidal response without significantly altering the overall stellar structure. Compared to the BM configurations, the CFL EoS sequences predict larger tidal deformabilities for the same stellar mass, indicating that the CFL EoS results in less compact stellar configurations. As shown in Figure \ref{fig:TidalBM}, the curves for different bag constants remain largely indistinguishable over the observed mass range, becoming distinguishable only near the maximum mass limit, as shown in Table \ref{tab:CFLLambda}.

Overall, both the BM and CFL EoS exhibit similar qualitative responses to the inclusion of $\DM$. Increasing the bag constant or introducing a $\DM$ component decreases the tidal deformability, with fermionic $\DM$ producing a slightly greater suppression than bosonic $\DM$. The predicted values of tidal deformability for the canonical, GW170817, and GW190425 configurations remain consistent with current gravitational-wave constraints, suggesting that the considered two-fluid stellar systems with $f_\x=5\%$ are compatible with existing observations.

\subsection{Universal Rotational-Tidal Relations}

\begin{figure}[!htp]
    \centering
   \includegraphics[width=8.7cm,height=9cm]{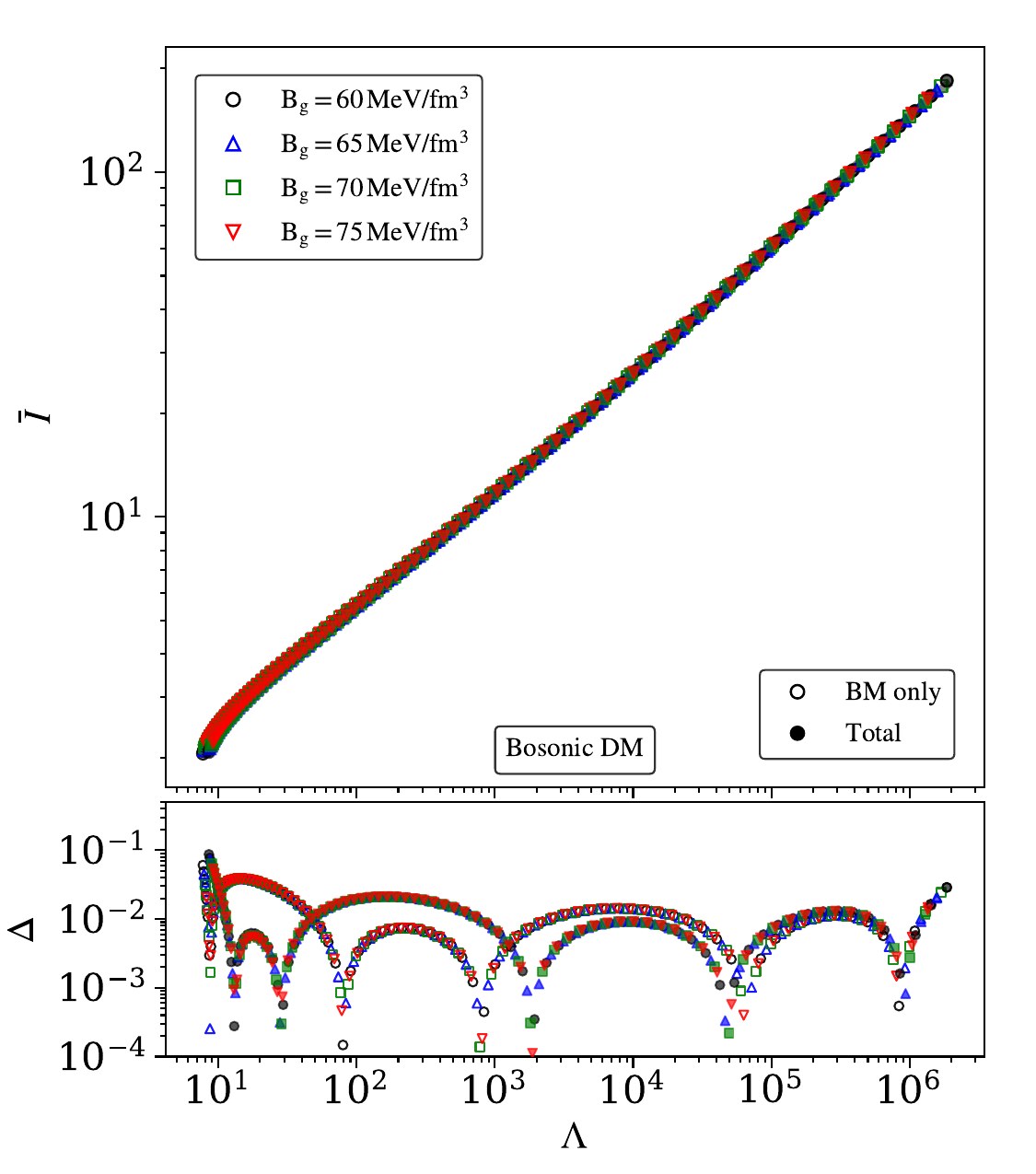}
    \vspace{0.2cm}
    \includegraphics[width=8.7cm,height=9cm]{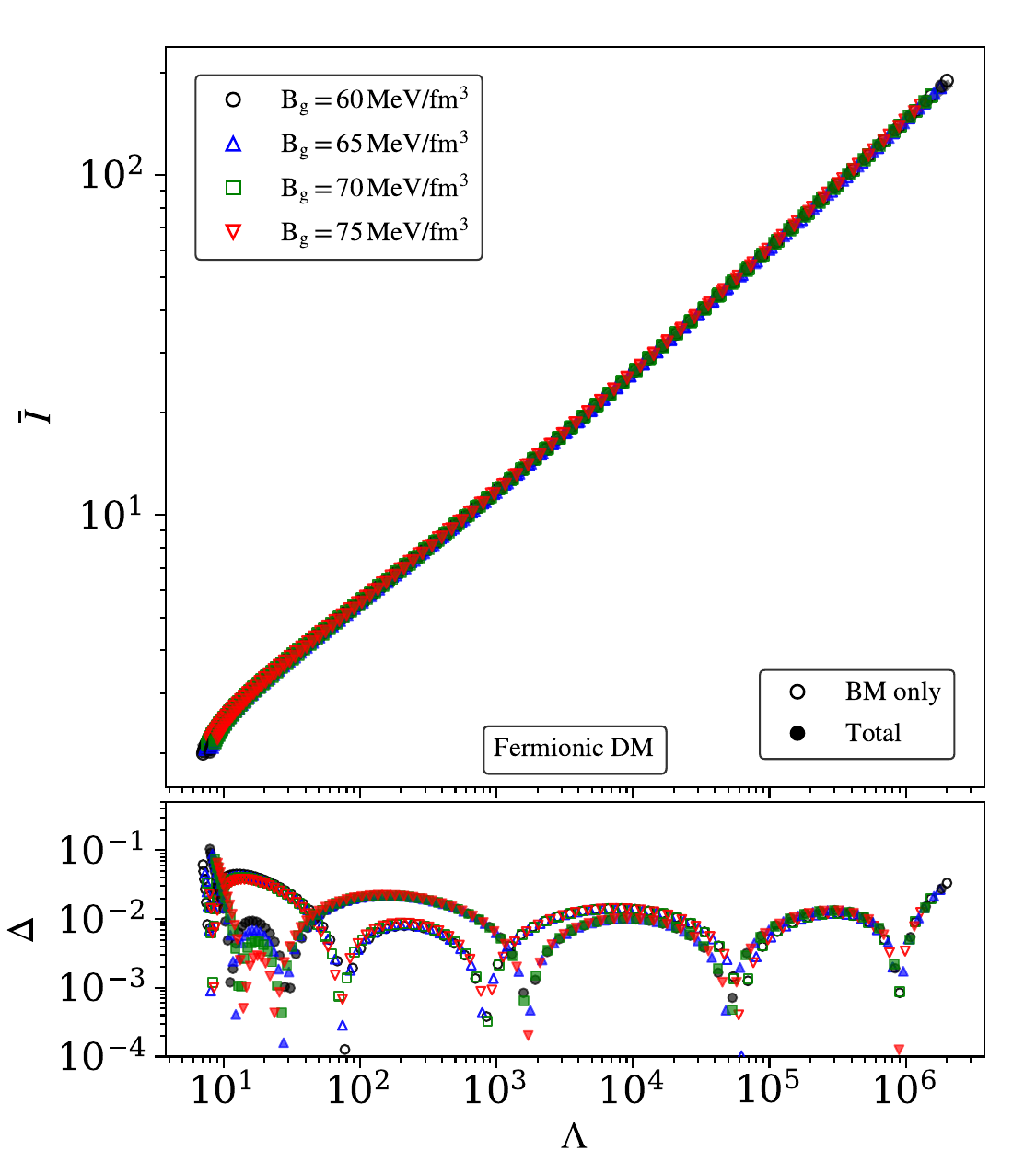}
    \caption{\justifying Universal $\I-\Lambda$ relations for quark star described by BM EoS admixed bosonic DM (upper-panel) and fermionic DM (lower-panel) for same configurations. The corresponding relative deviations, $\Delta$ from the fitted universal relation are shown in the lower subpanels. Open and filled markers denote configurations without and with DM, respectively.}
    \label{fig:IbarTidalBM}
\end{figure}

\begin{figure}[!htp]
    \centering
   \includegraphics[width=8.7cm,height=9cm]{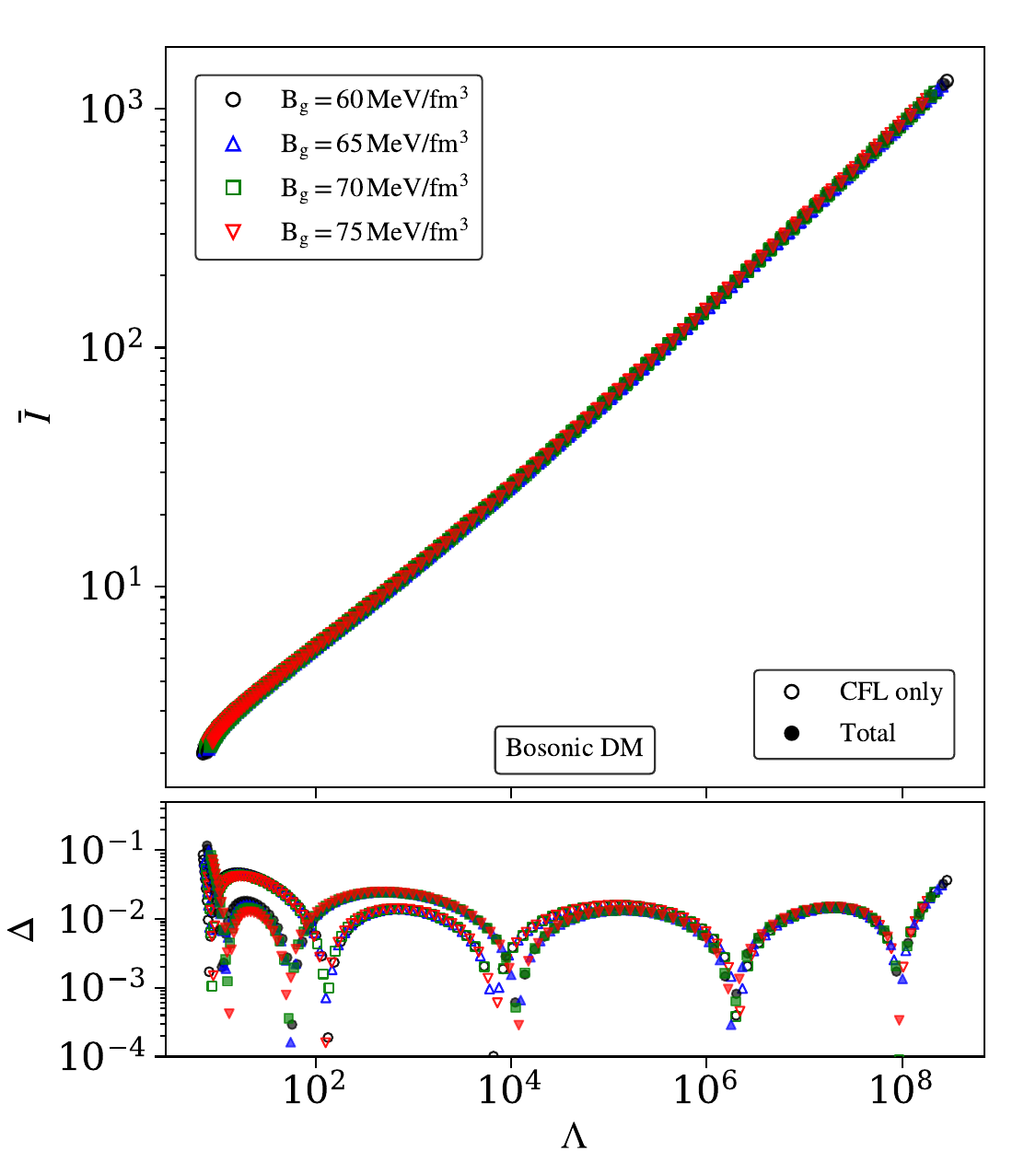}
    \vspace{0.2cm}
    \includegraphics[width=8.7cm,height=9cm]{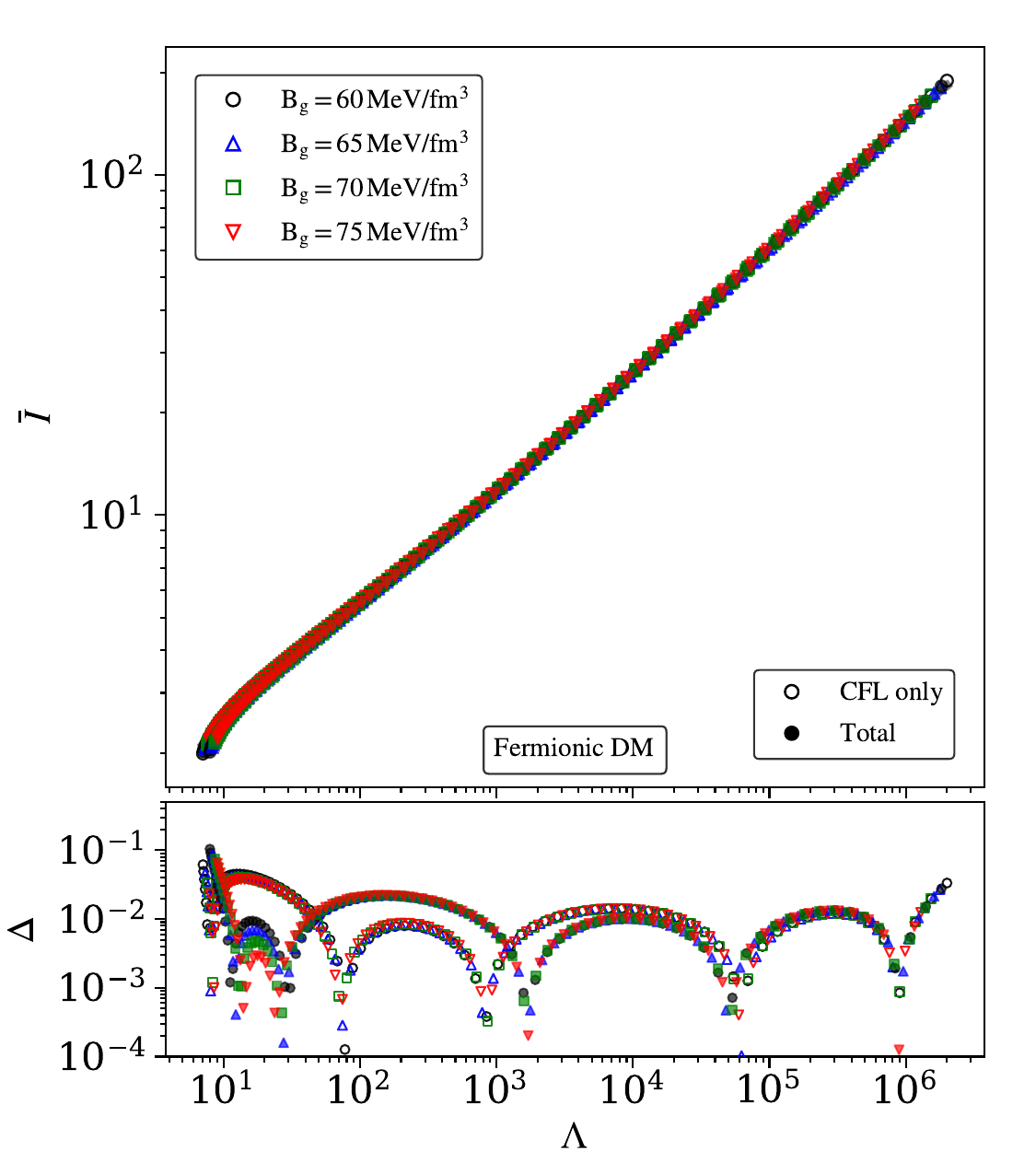}
    \caption{\justifying Universal $\I-\Lambda$ relations for quark stars described by CFL EoS admixed bosonic DM (upper-panel) and fermionic DM (lower-panel) for the same configurations. The corresponding relative deviations, $\Delta$ from the fitted universal relation are shown in the lower subpanels. Open and filled markers denote configurations without and with DM, respectively.}
    
    \label{fig:IbarTidalCFL}
\end{figure}

Finally, with the same configurations, we have generated the universal $\I$--Love relations, along with the relative deviation $\Delta$, for each model. Figs. \ref{fig:IbarTidalBM} and \ref{fig:IbarTidalCFL} present the universal $\I$--Love  relation for the BM and CFL EoS, respectively, each admixed with bosonic (upper panels) and fermionic DM (lower panels) for different values of $\B_g$. In both figures, open markers represent the OM configuration only, while solid markers represent the total configuration with DM fraction $f_\x = 5\%$. It is evident that the numerical curves corresponding to various values of $\B_g$ nearly coincide across the full range of $\Lambda$, indicating that, for both the BM and CFL EoS, the $\I$--Love relation is largely unaffected by the choice of bag constant. Moreover, introducing either bosonic or fermionic DM leads to only a very small modification of the universal relations, showing that incorporating a DM EoS does not substantially affect their universality.

%%%%%%%%%%%%%%%%%%%%%%%%%%%%%%%%%%%%%%%%%%%%%%
To characterize the variability, we additionally show lower sub-panels that depict the relative deviation between the empirical fit and the numerical data. The numerical values of the coefficients $\alpha_n$ for each model are listed in Table \ref{tab:ILoveCoeffs}. For the OM only configuration, the relative deviation peaks at $\sim 10\%$ in the low $\Lambda$ region before tapering down to just a few percentage points. A similar trend is observed for the total configuration (admixed with the DM EoS at the DM fraction $ f_\x=5\%$). However, the residuals exhibit slightly larger spread, particularly at small values of $\Lambda$. Despite this, the deviations remain relatively minor, suggesting that the $\bar{\I}-\Lambda$ universal relation is resilient against changes in the bag constant and the addition of $f_\x=5\%$, underscoring the persistence of the universality in the two-fluid stellar system configurations.

\section{CONCLUSION}\label{Sec:V}
We examined a fully relativistic stellar system employing a two-fluid framework, wherein OM and DM interact exclusively via gravity. We explored two EoS for OM, specifically the BM and CFL models, combined with two EoS for DM, namely bosonic and fermionic. Considering each fluid component as independently conserved, we formulate the coupled TOV equations of the two-fluid system and obtain the $\M-\R$ relations for the various values of the $\B_g$. Our analysis demonstrates that for each value of the $\B_g$, the inclusion of DM with a fixed $f_\x=5\%$ can significantly modify the properties of the stellar system.

The comparison between the bosonic and fermionic DM indicates that the bosonic DM generally supports configurations with slightly higher maximum masses than the corresponding fermionic DM case for both the BM and CFL EoS. The difference arises from the distinct pressure contributions of each DM EoS, leading to measurable changes in the stellar radius and maximum last stable star. Furthermore,  varying the bag constant $\B_g$ also alter the stiffness of the OM EoS and, as expected, increasing $\B_g$ softens the EoS, resulting in lesser masses and relatively small radii. To assess the astrophysical viability of the proposed models, we compared the equilibrium sequences of DM admixed quark star stellar systems with the currently available observational constraints. We observed that, for the adopted $f_\x$ and the considered ranges of the $\B_g$, both DM admixed configurations remain compatible with massive pulsars such as PSR J0740$+$6620, PSR J0952$-$0607, PSR J0348$+$0432, PSR J0437$-$4715 and PSR J0030$+$0451 and low-mass stellar remnants such as HESS J1731$-$347. Furthermore, the configurations also overlap with the parameter space favoured by the GW observations of binary NS mergers, particularly GW190425 and GW170817. Throughout the parameter space considered, all equilibrium models satisfy the Buchdahl compactness bound, the BH's limit and the causality limit, demonstrating that the inclusion of a modest DM component does not compromise the astrophysical consistency of the stellar models.

We further extended our analysis to a slowly rotating stellar system within the Hartle--Thorne perturbative formalism by retaining terms linear only in the stellar angular velocity. Assuming the $f_{_\Omega}=1$, we derive the corresponding frame dragging equation together with the generalized TOV equations and calculate the moment of inertia and mass relations for different models. For both the BM and CFL configurations, similar to $\M-\R$ relations, an increase in the $\B_g$ leads to lower maximum masses and correspondingly smaller maximum moments of inertia. Furthermore, for a fixed bag constant, the fermionic DM admixed configurations generally exhibit slightly larger maximum masses and moments of inertia than the corresponding bosonic DM models, although the overall behaviour of the $\I-\M$ relation remains qualitatively similar. Compared with the BM EoS, the CFL EoS generally yields larger moments of inertia. Moreover, the provided configurations are align with the existing observational constraints, supporting the astrophysical viability of both bosonic and fermionic DM admixed quark star stellar systems.

With the recent advances in multimessenger astronomy, we also investigated the tidal response by calculating the corresponding dimensionless tidal deformability for all the considered models. The resulting $\Lambda-\M$ shows that DM modifies the tidal response by altering by small amount of stellar compactness and internal structure, thereby shifting the deformability curves relative to purely OM configurations. For both the fermionic and the bosonic DM cases, the $\Lambda-\M$ curve closely overlaps with those corresponding to OM configurations, particularly in the low-mass regime, with only modest deviation at the higher mass. Furthermore, all DM admixed models remain consistent with observational constraints from GW events such as GW190425 and GW170817, as well as the canonical NS limits, supporting their astrophysical viability.

Finally, we provide the analysis of the universal $\I$--Love relations for the same combinations considered in this work. Despite the structural modifications introduced by the DM component, the calculated configurations continue to exhibit an approximately EoS behaviour. The numerical results for different values of $\B_g$ almost completely overlap over the entire range of $\Lambda$, demonstrating that the $\I$--Love relation is largely insensitive to the bag constant for both the BM and CFL EoS. Higher relative deviations ($\sim 10\%$) are observed depending on lower $\Lambda$ for all the values of the bag constant. However, these deviations remain relatively small for the higher $\Lambda$ ($>10^1$), demonstrating the remarkable robustness of the universal relation even in the presence of a modest DM admixture. This finding indicates that the $\bar{\I}$--$\Lambda$ relation continues to provide a valuable diagnostic for probing the internal composition of compact stars and assessing possible DM effects.

We conclude that our analysis demonstrates that the DM admixed quark stars constitute a promising laboratory for exploring the interplay between dense matter physics and particle DM.  The compatibility of both bosonic and fermionic DM admixed configurations with the current observational constraints on the masses, radii, moments of inertia, and tidal deformabilities demonstrates their astrophysical viability for the adopted ranges of bag constant and also DM fraction, $f_\x=5\%$. Future multi-messenger observations, combining increasingly precise measurements from radio pulsars, NICER, and next-generation GW detectors, will offer the opportunity not only to further constrain the EoS of dense matter but also to distinguish among different DM scenarios through their subtle yet observable signatures in the properties of compact stellar systems.

\section*{Acknowledgement}
 The Author SKM is thankful for the continuous support and encouragement from the administration of the University of Nizwa for this research work. NP acknowledges the financial support provided by the University Grants Commission (UGC Ref. No.: 231620138510) through the Senior Research Fellowship to carry out the research work.

\section*{References}
\bibliographystyle{Gen.bst}
\bibliography{Reference}
\end{document}